\documentclass{aastex631}

\usepackage{amsmath,bm}
\usepackage{longtable}
\usepackage{natbib}
\usepackage[T1]{fontenc}
\usepackage{newtxtext,newtxmath}
\DeclareRobustCommand{\VAN}[3]{#2}
\let\VANthebibliography\thebibliography
\def\thebibliography{\DeclareRobustCommand{\VAN}[3]{##3}\VANthebibliography}

\usepackage{graphicx}	% Including figure files
\usepackage{bm}
\usepackage{amsmath}	% Advanced maths commands
\usepackage{booktabs}
\usepackage{gensymb}
\begin{document}

%%%%%%%%%%%%%%%%%%% TITLE PAGE %%%%%%%%%%%%%%%%%%%

% Title of the paper, and the short title which is used in the headers.
% Keep the title short and informative.
\title[NGC 5728]{Deep \textit{Chandra} Observations of NGC 5728: Morphology and Spectral Properties of the Extended X-ray Emission}

% The list of authors, and the short list which is used in the headers.
% If you need two or more lines of authors, add an extra line using \newauthor
\correspondingauthor{Anna Trindade Falcao}
\email{anna.trindade\_falcao@cfa.harvard.edu}

\author{Anna Trindade Falcao}
\affiliation{Harvard-Smithsonian Center for Astrophysics, \\
60 Garden St., Cambridge, MA 02138, USA}

\author{G. Fabbiano}
\affiliation{Harvard-Smithsonian Center for Astrophysics, \\
60 Garden St., Cambridge, MA 02138, USA}

\author{M. Elvis}
\affiliation{Harvard-Smithsonian Center for Astrophysics, \\
60 Garden St., Cambridge, MA 02138, USA}

\author{A. Paggi}
\affiliation{Dipartimento di Fisica, Universita' degli Studi di Torino, via Pietro Giuria 1, \\
I-10125, Torino, Italy}
\affiliation{Istituto Nazionale di Fisica Nucleare, Sezione di Torino, via Pietro Giuria 1, \\
10125, Torino, Italy}

\author{W. P. Maksym}
\affiliation{Harvard-Smithsonian Center for Astrophysics, \\
60 Garden St., Cambridge, MA 02138, USA}




\begin{abstract}
 Recent deep \textit{Chandra} observations of nearby Compton thick (CT) AGN have produced surprising results, uncovering extended emission not only in the soft X-rays but in the hard emission ($>$3 keV), challenging the long-held belief that the characteristic hard X-ray continuum and fluorescent Fe K lines are associated with the torus in the standard picture of AGN. In this work, we present the analysis of our deep ($\sim$261 ks) X-ray \textit{Chandra} ACIS-S observations of NGC 5728, a nearby ($z$=0.00932) CT AGN. We find that the diffuse emission is more extended at lower energies, in the bicone direction out to $\sim$ 2 kpc radially, but also significantly extended in the direction of the cross-cone, out to $\sim$1.4 kpc. Our results suggest that the ratio of detected photons in the cross-cone to the bicone region is $\sim$ 16\%, below 3 keV, decreasing to 5\% for energies 3-6 keV. The nuclear spectrum suggests a low photoionization phase mixed with a more ionized gas component, while the bicone and cross-cone spectra are dominated by a mix of photoionization and shocked gas emission. A mixture of thermal and photoionization models to fit the spectra indicates the presence of complex gas interactions, consistent with previous observations of other CT AGN (e.g., ESO 428–G014).

\end{abstract}


%%%%%%%%%%%%%%%%% BODY OF PAPER %%%%%%%%%%%%%%%%%%

\section{Introduction}
The interaction between the radiation released by accreting supermassive black holes (SMBHs) in active galactic nuclei (AGN) and the interstellar medium (ISM) is thought to play a crucial role in the evolution of their host galaxies \citep[e.g.][]{silk1998a, dimatteo2005a}. The radiation emitted by the SMBH exists in the form of jets, winds, and outflows and can generate efficient feedback when such interaction initiates a feedback loop that self-regulates star formation and SMBH growth.\par 

In this paper, we analyze the deep \textit{Chandra} observations of the nuclear region of the nearby ($z$=0.00932) active barred spiral, type 2 galaxy NGC 5728, cited as the paradigm of Type 2 AGN with ionization cones by \citet{urry1995a} in their review paper on the AGN unification model. At a distance of approximately 41 Mpc \citep[][scale=200 pc/arcsec]{mould2000a}, NGC 5728 is a CT AGN with a heavily obscured (log NH=24.3, \citealt{koss2017a}) and complex nucleus. In the cross-cone direction, the circumnuclear area exhibits a strong star-forming ring, evident in H~II, while it also shows a prominent ionization bicone in the perpendicular direction, as well as a one-sided radio jet \citep{wilson1993a, durre2018a, durre2019a}. In recent studies, \citet{durre2018a, durre2019a} analyzed the spatially-resolved kinematics of the ionized gas in the central region of this galaxy and found that significant amounts of gas ($\sim$ 38 M$_{\odot}$ yr$^{-1}$) are being cleared from this area due to the impact of powerful AGN-driven outflows.\par

Based on the first $\sim$19 ks archival \textit{Chandra} ACIS-S image, the X-ray emission in the bicone direction appears extended in the soft band ($<$ 3 keV), a phenomenon that has also been observed in other AGN \citep[e.g.,][]{wang2011a, fabbiano2018a}. Similar to the case of the CT AGN ESO 428-G014 \citep{fabbiano2018a}, extended X-ray emission in the 3-6 keV band is also suggested in the bicone direction. A strong point-like source, absorbed in the soft band, arises at these higher energies. The X-ray spectrum is a featureless continuum in the 3-6 keV spectral range, and the nuclear source is associated with the strong 6.4 keV Fe-K$\alpha$ line emission in the archival data.\par 

With the cumulative \textit{Chandra} exposure of $\sim$261 ks, we reexamined the emission in NGC 5728, aiming at determining the detailed morphological and spectral properties of the extended X-ray emission. Because NGC 5728 is a CT AGN, the X-ray luminosity originating in the nuclear source is heavily attenuated by the torus, and it is not piled up in ACIS-S, allowing us to investigate the circumnuclear region to the smallest sub-arcsecond radii permitted by \textit{Chandra} resolution. The scientific questions we hope to answer for this galaxy all revolve around improving our understanding of AGN-galaxy interaction and feedback. While feedback is an essential component of galaxy evolution, it is only now, with high-quality multi-wavelength data, that a direct observational understanding of the phenomena at play can be obtained. \par 

In this work, we analyze the extended emission in NGC 5728 in the full (0.3-7 keV) energy band observable with \textit{Chandra}. In Section \ref{sec:data_analysis}, we describe the observations, reduction and analysis of the data, and the alignment procedures applied to these observations. In Section \ref{sec:images}, we describe the methods used for the imaging analysis, including modeling of the \textit{Chandra} PSF, and surface brightness radial profiles. Section \ref{sec:spec_analysis} describes the methods used for the spectral analysis, including the classes of models used to fit the extracted spectra. We then report the final results of our imaging and spectral analysis, both for the extended and the nuclear emission, in Section \ref{sec:results}. Finally, in Section \ref{sec:discussion}, we discuss our findings and present our conclusions in Section \ref{sec:conclusions}.

\section{Observations and Data Analysis}
\label{sec:data_analysis}

NGC 5728 was observed with \textit{Chandra} ACIS-S eleven times, with a total exposure time of 260.8 ks, as shown in Table \ref{tab:obs}. We retrieved the data from the \textit{Chandra} archive\footnote{https://cda.harvard.edu/chaser/}, and used \textsc{ciao} 4.14\footnote{http://cxc.harvard.edu/ciao} to reduce the observations. We reprocessed the data with the standard \textsc{chandra\_reprocess} script \citep{fruscione2006a}, taking advantage of the sub-pixel capabilities of the ACIS detector due to the spacecraft's dither; individual observations were astrometrically aligned to the longest observation (ObsID 22582). \par 

\begin{table}[htb!]
\footnotesize
\begin{center}
\caption{\textbf{Details of the used \textit{Chandra} ACIS-S observations and applied offsets.}}
\label{tab:obs} 
\begin{tabular}{ccccc}
\multicolumn{5}{c}{}\\
\hline
\multicolumn{1}{c}{Obs ID}
&\multicolumn{1}{c}{Date}
&\multicolumn{1}{c}{Exp Time (ks)}
&\multicolumn{1}{c}{P.I.}
&\multicolumn{1}{c}{Applied Offsets (sky pixels)}

\\
\hline
\hline
4077 & 2003-06-27 & 18.73 & Kraemer & $\Delta$x=-0.03 \\
 &  &  &  &  $\Delta$y=0.77 \\
 \hline
22582 & 2019-12-29 & 49.42 & Fabbiano & ASTROMETRIC \\
& & & & REFERENCE \\
\hline
22583 & 2020-05-22 & 29.68 & Fabbiano & $\Delta$x=-0.91 \\
& & & & $\Delta$y=-0.51 \\
\hline
23041 & 2020-04-18 & 13.89 & Fabbiano & $\Delta$x=0.24 \\
& & & & $\Delta$y=1.08 \\
\hline
23042 & 2021-04-12 & 25.73 & Fabbiano & $\Delta$x=1.06\\
& & & & $\Delta$y=-0.06 \\
\hline
23043 & 2020-05-12 & 19.81 & Fabbiano & $\Delta$x=-0.64 \\
& & & & $\Delta$y=-0.17 \\
\hline
23221 & 2020-04-16 & 15.86 & Fabbiano & $\Delta$x=0.08\\
& & & & $\Delta$y=0.96\\
\hline
23249 & 2020-05-14 & 29.68 & Fabbiano & $\Delta$x= -1.09\\
& & & & $\Delta$y=-0.44 \\
\hline
23254 & 2020-05-23 & 19.82 & Fabbiano & $\Delta$x=-1.08\\
& & & & $\Delta$y=1.09\\
\hline
25006 & 2021-04-12 & 21.79 & Fabbiano & $\Delta$x=1.34\\
& & & & $\Delta$y=-0.59\\
\hline
25007 & 2021-04-16 & 16.37 & Fabbiano & $\Delta$x=1.00\\
& & & & $\Delta$y=0.42\\

\hline
\hline
\end{tabular}
\end{center}
\end{table}

To build the most accurate merged data set, we examined two different ways of aligning the images from different observations. The first approach was to align the observations using the centroids of the images in the 6-7 keV energy band, which has the advantage of being dominated by the nuclear position in CT AGN. To determine the centroid of the nuclear source for each individual observation, we built images in the 6-7 keV energy band, with subpixel scale of 1/8 of the native ACIS pixel (smoothed with a 2D Gaussian kernel with $\sigma$=2), selecting the brightest pixel in the output image. We used \textsc{dmstat} to calculate the centroid of the count distribution within a circle of radius $r$=0.2$''$, centered on the brightest pixel of the smoothed image. We then used the tool \textsc{wcs\_update} to apply the offsets, with shifts input to the \textsc{deltax} and \textsc{deltay} parameters in units of sky pixels (Table \ref{tab:obs}). Finally, we used the new aspect solution to re-calculate the World Coordinate System (WCS) events positions in the original event file with the \textsc{reproject\_obs} tool.\par

We confirmed the accuracy of our results by evaluating off-nuclear point sources in the field of view ($<$8$''$ away from the active nucleus). We used \textsc{wavdetect}, with $\sqrt{2}$ scales\footnote{With 1.0, 1.4, 2.0, 2.8, 4.0, 5.7, 8.0, 11.3, 16.0, used to get a more extensive run.} adopted as the wavelet parameter and a significance threshold of 10$^{-6}$ false detections per pixel. By comparing the offsets of the detected point sources centroids in the single observations relative to the final merged image (in the 0.3-7 keV band), we found that this method yields nuclear positions consistent with those found by aligning the image centroids in the 6-7 keV energy band.\par 

%\textbf{We compared the FWHM of the Gaussian components modeling the nuclear source in the merged images in the 6-7 keV band to select the most precise merging method. We used a one-dimension Gaussian function, i.e., the function \textit{gauss1d} in \textit{Sherpa}\footnote{https://cxc.harvard.edu/sherpa/}, to fit the nuclear source, and extracted the FWHM of the best-fit Gaussian models along the major axis. The FWHM in the 6-7 keV merged image from the nuclear alignment method is XXX native pixels, which is XXX times the size of the pre-flight PSF in the same energy band (FWHM=XXX native pixels). Fitting the nuclear source in the off-nuclear point-source merged image at 6-7 keV results in a slightly larger FWHM (XXX native pixels). Although these values are consistent at 1$\sigma$, the method of aligning the observations using the centroids in the 6-7 keV energy band produces the most precise merged image.}\par 

\begin{figure}[htb!]
  \centering
  %\begin{adjustwidth}{-.05in}{-0in}
  \includegraphics[width=18cm]{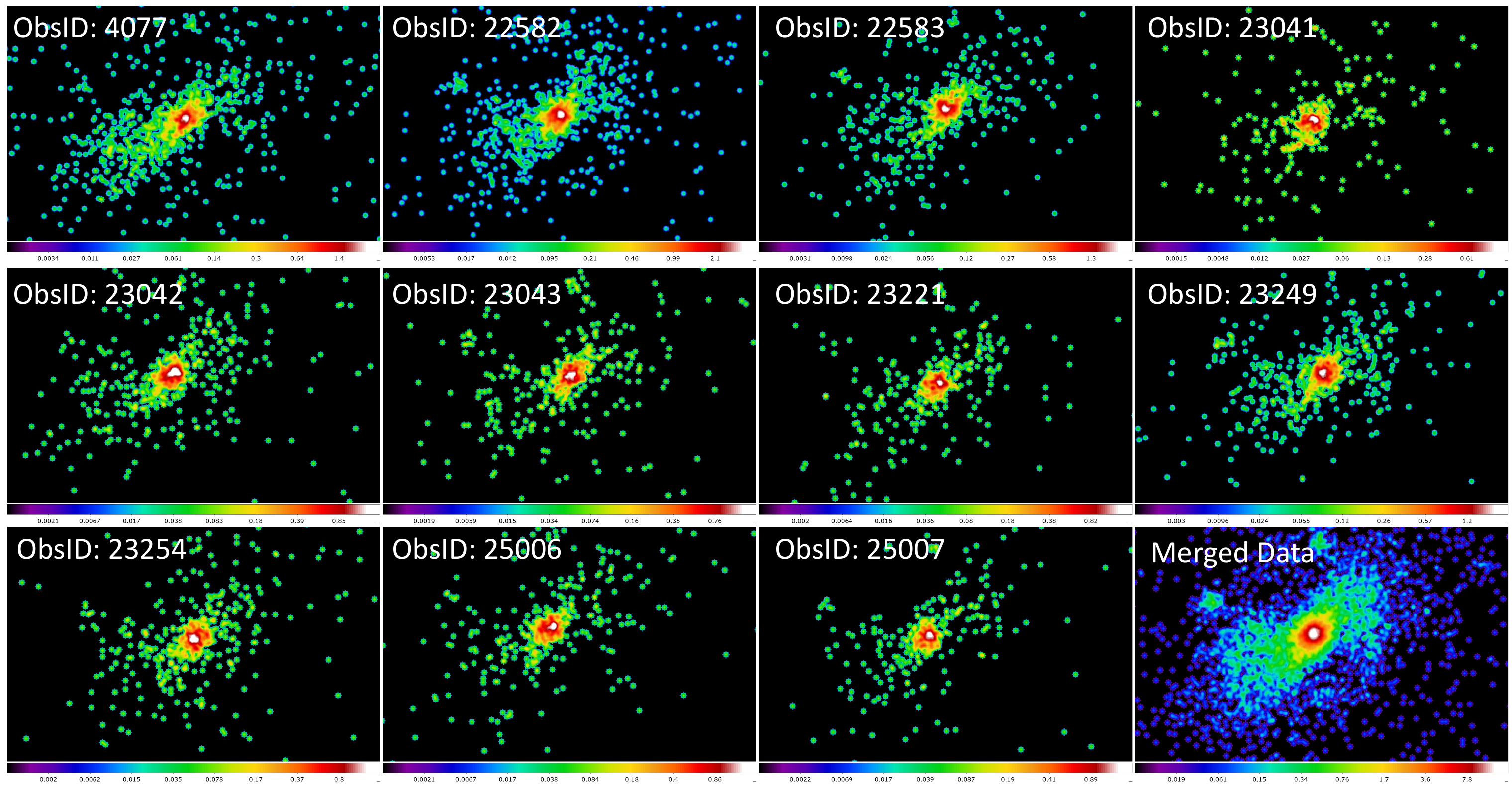}
\caption{1/8 subpixel 0.3-7 keV images of all observations plus the final merged image (bottom right panel), which was obtained by aligning all eleven images. N is up, and E is to the left. All panels are 20$''$ $\times$ 15$''$ in size. Images were smoothed with 4 pixel Gaussian.}
 \label{fig:obs} 
 \end{figure}

To identify fainter emission features, we created a merged event file using \textsc{merged\_obs}. The reprojected images of individual observations and the merged image are shown in Fig. \ref{fig:obs}. These are 20$''$ $\times$ 15$''$, 1/8 subpixel images, smoothed with 4 pix Gaussian, in the 0.3-7 keV energy band.\par

\newpage 

\section{Image Analysis}
\label{sec:images}

\subsection{\textit{Chandra} PSF Modeling}

Following \textit{Chandra} science threads\footnote{https://cxc.cfa.harvard.edu/ciao/threads/}, we simulated the point spread function (PSF) of all individual observations listed in Table \ref{tab:obs}. To do so, we used \textsc{ChaRT}\footnote{https://cxc.harvard.edu/ciao/PSFs/chart2/}, taking into account the corresponding individual source spectrum (extracted with \textsc{specextract}\footnote{https://cxc.cfa.harvard.edu/ciao/threads/marx\_sim/}), aspect solution, and off-axis angle as inputs for the simulations. We then projected the PSF rays onto the detector plane with \textsc{marx}, taking into account the nominal position of the detector during the observation, the observation date, and the exposure time. We also set the appropriate telescope and detector configuration and corrected for science instrument module (SIM) offset. To avoid any effects due to pixelization, we set conservatively \textsc{AspectBlur}=0.2\footnote{https://cxc.cfa.harvard.edu/ciao/why/aspectblur.html}, as recommended in \textsc{marx} for ACIS-S observations. Each \textsc{ChaRT} simulation was individually projected and made into event files, which were then combined into a single event file using \textsc{dmmerge}. We used ObsID 22582 as our astrometric reference to merge the PSF models, and the final PSF event file was normalized to the source counts within a circle of 0.5$''$ radius. \par 

\subsection{Probing the Extended X-ray Emission}

\begin{figure}[htb!]
  \centering
  %\begin{adjustwidth}{-.05in}{-0in}
 \includegraphics[width=18cm]{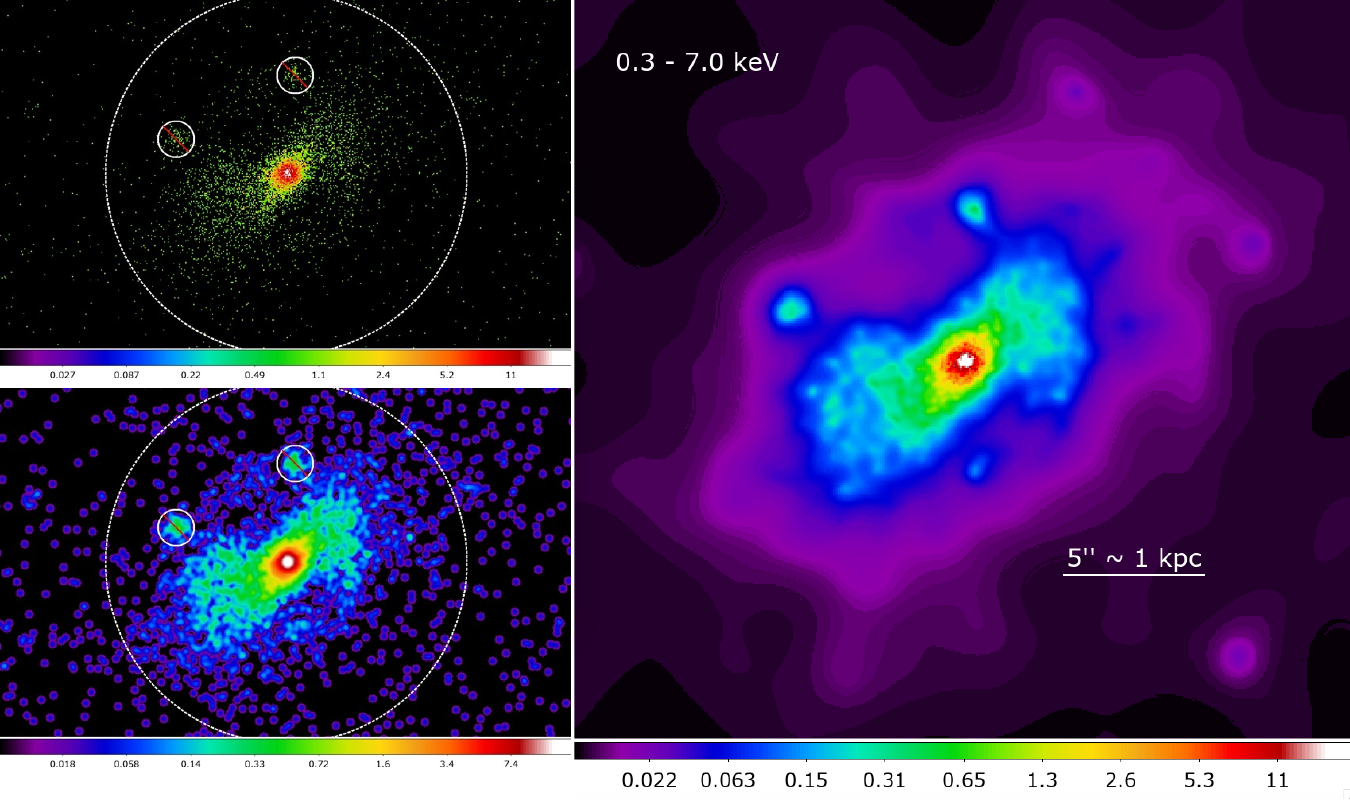}
\caption{Emission in the full-band (0.3-7 keV). \textit{\textbf{Top-Left:}} "raw" 1/8 subpixel merged image. \textit{\textbf{Bottom-Left:}} 1/8 subpixel merged image smoothed with 4 pix Gaussian. \textit{\textbf{Right:}} merged image adaptively smoothed with Gaussian (0.5-30 pix scales, 10 counts under the kernel, 60 iterations). The regions of non-nuclear point sources excluded from the spectral analysis and surface brightness profiles are marked. Dashed circles have a radius of 10$''$ ($\sim$ 2 kpc).}
 \label{fig:merged} 
 \end{figure}

The final full-band image of NGC 5728 is shown in more detail in Fig. \ref{fig:merged}. The upper-left panel shows the "raw" 1/8 subpixel merged image, while the lower-left panel shows the 1/8 subpixel merged image smoothed with 4 pix Gaussian. The right panel shows the merged image adaptively smoothed with Gaussian (0.5-30 pix scales, 10 counts under the kernel, 60 iterations). This image shows a prominent point-like nuclear source and fainter X-ray emission extended $\sim$10$''$ ($\sim$2 kpc) radially from the nucleus, in the SE-NW direction, which is also the direction of the ionization bicone \citep{durre2018a}. \par 

% \begin{figure} [h!]
%    \centering
%    \includegraphics[width=18cm]{comp_subtraction.png}
%    \caption{\textbf{Left:} Merged \textit{Chandra} ACIS data (0.3- 7 keV) with subpixel binning (1/8 of native pixel size). \textbf{Middle:} Merged PSF simulations (0.3- 7 keV). \textbf{Right:} Merged data minus merged PSF simulations, providing a visualization of the extended X-ray emission morphology in this galaxy.}
%    \label{fig:psf_comp}
%\end{figure}

%We energy-filtered the 1/8 subpixel final image in six energy bands. As shown in the leftmost column in Figures \ref{fig:radial_1} and \ref{fig:radial_2}, the emission appears more extended at lower energies, although elongated diffuse surface brightness can be seen in the \textbf{SE-NW} direction at energies greater than 3 keV. This emission seems to be present but less prominent, at least up to 7 keV in the inner $\sim$ 5$''$ (1 kpc). \par 

\begin{figure*} [h!]
    \centering
    \includegraphics[width=16cm]{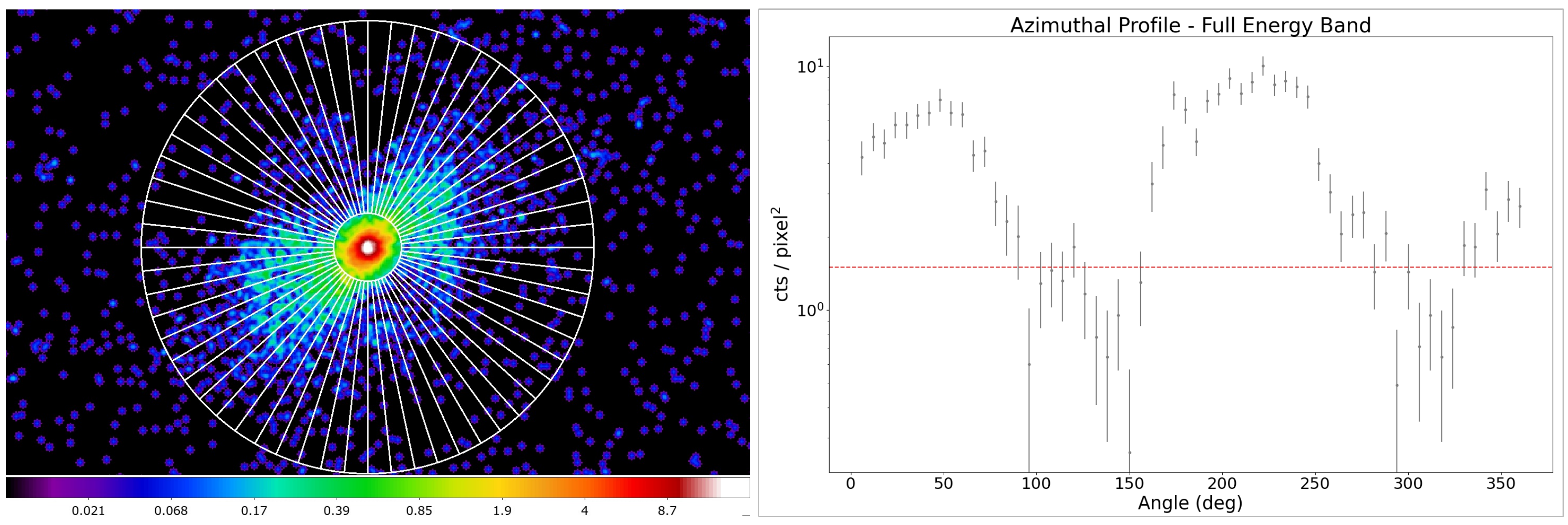}
    \caption{\textbf{Left:} \textit{Chandra} full energy band image with overlaying 1.5$''$-10$''$ annular extraction region used to extract the azimuthal profile from the merged data (60 angular bins of 6$\degree$. \textbf{Right:} Merged data azimuthal profile (in gray). The red horizontal line indicates 1.5 counts/pixel$^{2}$. }
    \label{fig:azimuthal}
\end{figure*}

 \begin{figure}
 \begin{minipage}[b]{0.3\textwidth}
  \includegraphics[width=18cm]{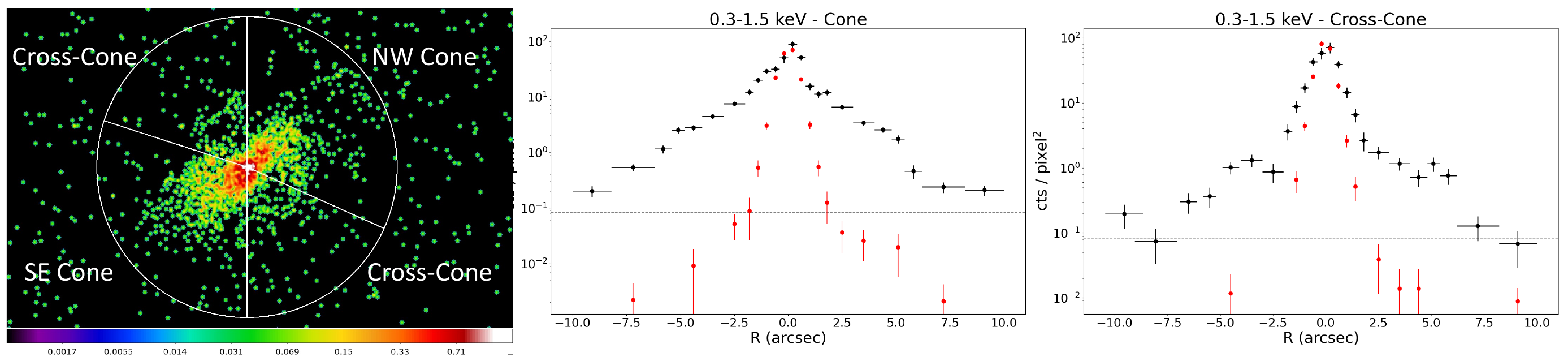}
 \end{minipage}\qquad \\
 
 \begin{minipage}[b]{0.3\textwidth}
  \includegraphics[width=18cm]{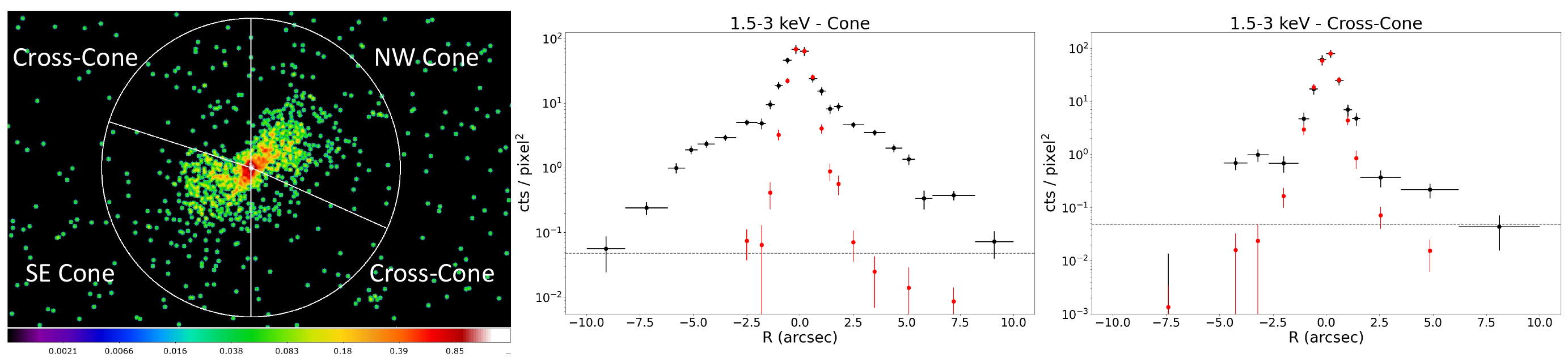}
 \end{minipage}\qquad \\
 
 \begin{minipage}[b]{0.3\textwidth}
  \includegraphics[width=18cm]{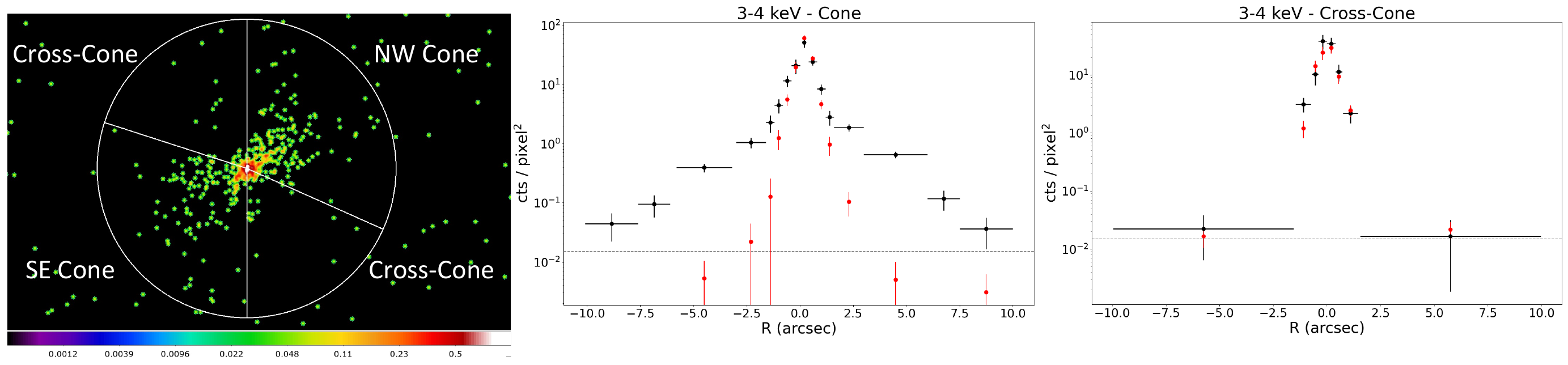}
 \end{minipage}\qquad \\

 \caption{Images (leftmost column), and background-subtracted surface brightness radial profiles in the bicone (middle column) and cross-cone (rightmost column) sectors of NGC 5728 in different energy bands (top panels: 0.3–1.5 keV; middle panels: 1.5–3 keV; bottom panels: 3–4 keV). The images were re-binned at 1/8 of native pixel and smoothed with Gaussian (setting the following parameters: radius = 4, sigma = 2, in pixels). The PSF in the same energy band was normalized to the source image in the central 0.5$''$ circle. The bin size was chosen to contain a minimum of 10 counts in the image data set (extraction regions with less than 10 counts were not plotted).The dashed gray horizontal line indicates the level of the field background for each plot.}
\label{fig:radial_1}
%\end{adjustwidth}
\end{figure}

\begin{figure}
\begin{minipage}[b]{0.3\textwidth}
  \includegraphics[width=18cm]{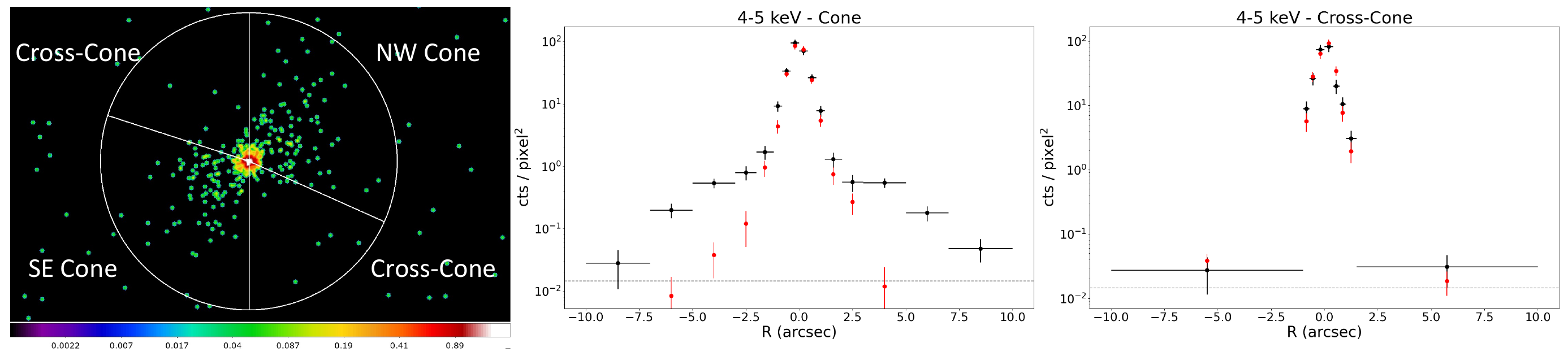}
 \end{minipage}\qquad \\ 
 
 \begin{minipage}[b]{0.3\textwidth}
  \includegraphics[width=18cm]{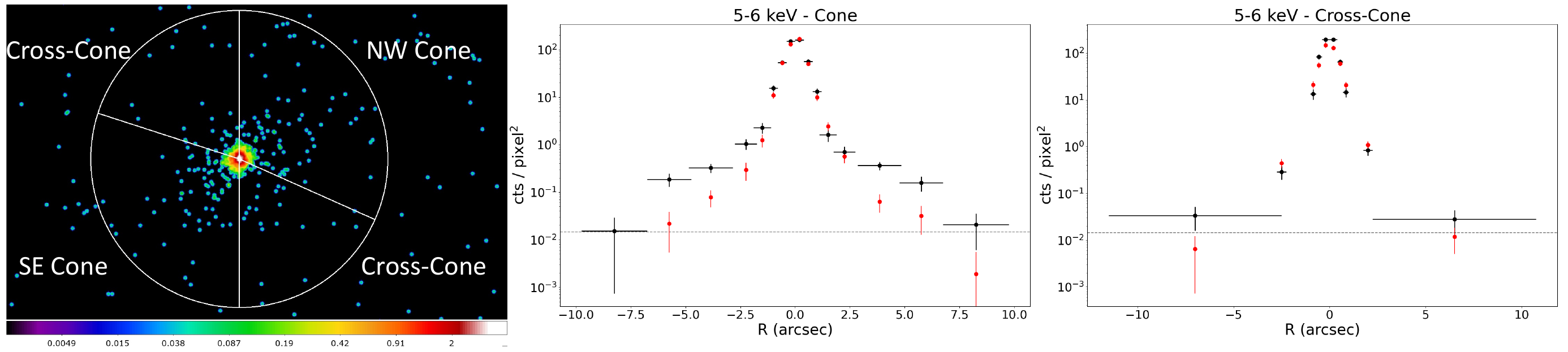}
 \end{minipage}\qquad \\ 
 
 \begin{minipage}[b]{0.3\textwidth}
  \includegraphics[width=18cm]{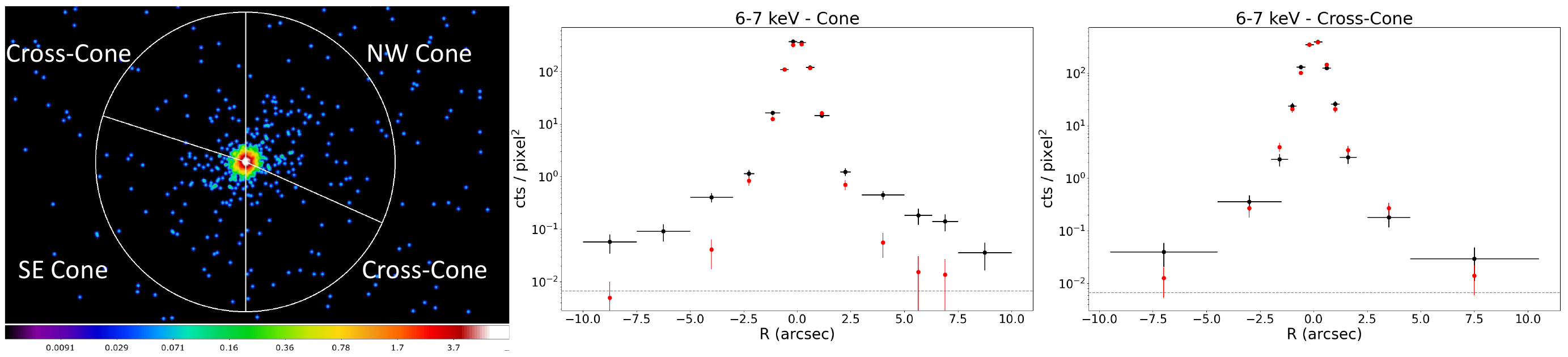}
 \end{minipage}\qquad \\ 
 
  \caption{Top row: 4–5 keV; middle row: 5–6 keV; bottom row: 6-7 keV. See the caption in Fig. \ref{fig:radial_1} for details.}
\label{fig:radial_2}
%\end{adjustwidth}
\end{figure}

\begin{figure}[htb!]

\begin{minipage}[b]{.4\textwidth}
  \includegraphics[width=18cm]{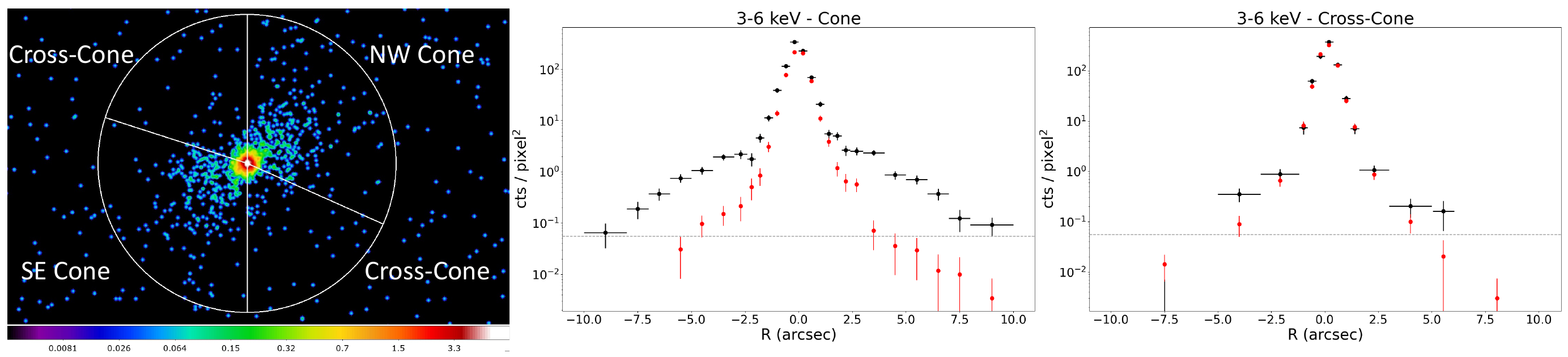}
 \end{minipage}\qquad\\
 \begin{minipage}[b]{.4\textwidth}
  \includegraphics[width=18cm]{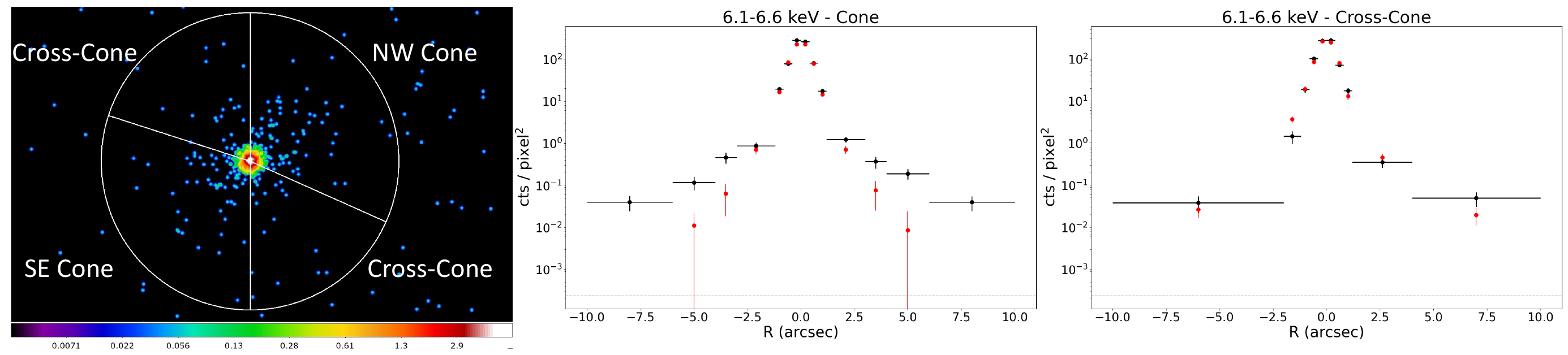}
 \end{minipage}\qquad \\

\caption{Same as in Fig. \ref{fig:radial_1}, for 3-6 keV (top) and the Fe K$\alpha$ band (bottom).}
 \label{fig:3-6kev} 
 \end{figure}

We sliced the final merged image in bicone and cross-cone sectors to investigate each region's circumnuclear gas morphology and extent separately. To optimally determine the angles of these regions, we produced a surface brightness azimuthal profile of the 0.3–7 keV merged image within the 1.5$''$–10$''$ annular region centered on the nucleus, with angular bins of 6$\degree$ (Fig. \ref{fig:azimuthal}). The profile obtained was fitted with two Gaussians plus a constant to account for the background level. Based on the Gaussian centers and widths, we selected the bicone angles to be: NW cone between 336$\degree$ and 90$\degree$ (measured counter-clockwise from north), and the SE cone between 162$\degree$ and 270$\degree$. Similarly, the cross-cone angles are defined as: NE cross-cone between 90$\degree$ and 162$\degree$, and SW cross-cone between 270$\degree$ and 336$\degree$, as shown in Fig. \ref{fig:radial_1}, \ref{fig:radial_2}, and \ref{fig:3-6kev} (leftmost column).  \par 

To quantify the magnitude and significance of the extended components in different energy bands, we energy-filtered the merged event file and the merged \textit{Chandra} PSF model, following \textit{Chandra} threads\footnote{https://cxc.cfa.harvard.edu/ciao/threads/radial\_profile/} to extracted the surface brightness profile of both data sets. We used \textsc{wavdetect} to identify off-nuclear point sources in the final merged image, excluding these from the radial profiles, as shown in Fig. \ref{fig:merged}. The rightmost columns in Fig. \ref{fig:radial_1}, \ref{fig:radial_2}, and \ref{fig:3-6kev} show a comparison between the radial surface brightness profiles of the merged image (in black) and merged model PSF (in red), in different energy bands in the cone and cross-cone directions. The data and PSF radial surface brightness profiles were extracted from 1/8 subpixel images, with bin sizes varying to contain a minimum of 10 counts per bin in the image data set. Because the model PSF was built to match the cumulative exposure time of the data, the noise observed in the radial profiles is consistent with statistical noise. The gray horizontal line indicates the level of the field background derived from the same image as the profiles. Since the profiles are background subtracted, points below the background level are valid data. Outside the plotted range, the profiles are dominated by noise. \par

\begin{table}[htb!]
\footnotesize
\begin{center}
\caption{\textbf{Excess Data Counts over Model PSF for 1.5$''$-8$''$ Conical Regions.}}
\label{tab:excess} 
\begin{tabular}{cccc}
\multicolumn{4}{c}{}\\
\hline
\multicolumn{1}{c}{}
&\multicolumn{1}{c}{NW Cone Sector}
& \multicolumn{1}{c}{SE Cone Sector}
&\multicolumn{1}{c}{Cross-Cone Sectors}
\\
Energy (keV) & --------------- & --------------- & --------------- \\
& Excess (error) & Excess (error) & Excess (error) \\
\hline
0.3-7.0  & 1189 (34) & 1216 (35) &  324 (18)\\
0.3-1.5  & 466 (22) & 605 (25) & 215 (15)\\
1.5-3.0  & 421 (21) & 376 (19) & 78 (9)\\
3.0-4.0  & 132 (11) & 83 (9) & 18 (4)\\
4.0-5.0  & 75 (9) & 67 (8) & 10 (3)\\
5.0-6.0  & 32 (6) & 42 (7) & 3 (2)\\
5.8-6.2  & 21 (5) & 25 (5) & 6 (3) \\
6.0-7.0  & 63 (7) & 43 (6) & - \\
6.1-6.6  & 43 (6) & 32 (6) & -\\
6.7-7.2   & 8 (3) & 16 (4) & 3 (2) \\
\hline
\end{tabular}
\end{center}
\end{table}

\newpage

 \section{Spectral Analysis}
\label{sec:spec_analysis}

\begin{figure}[htb!]
  \centering
  %\begin{adjustwidth}{-.05in}{-0in}
 \includegraphics[width=9cm]{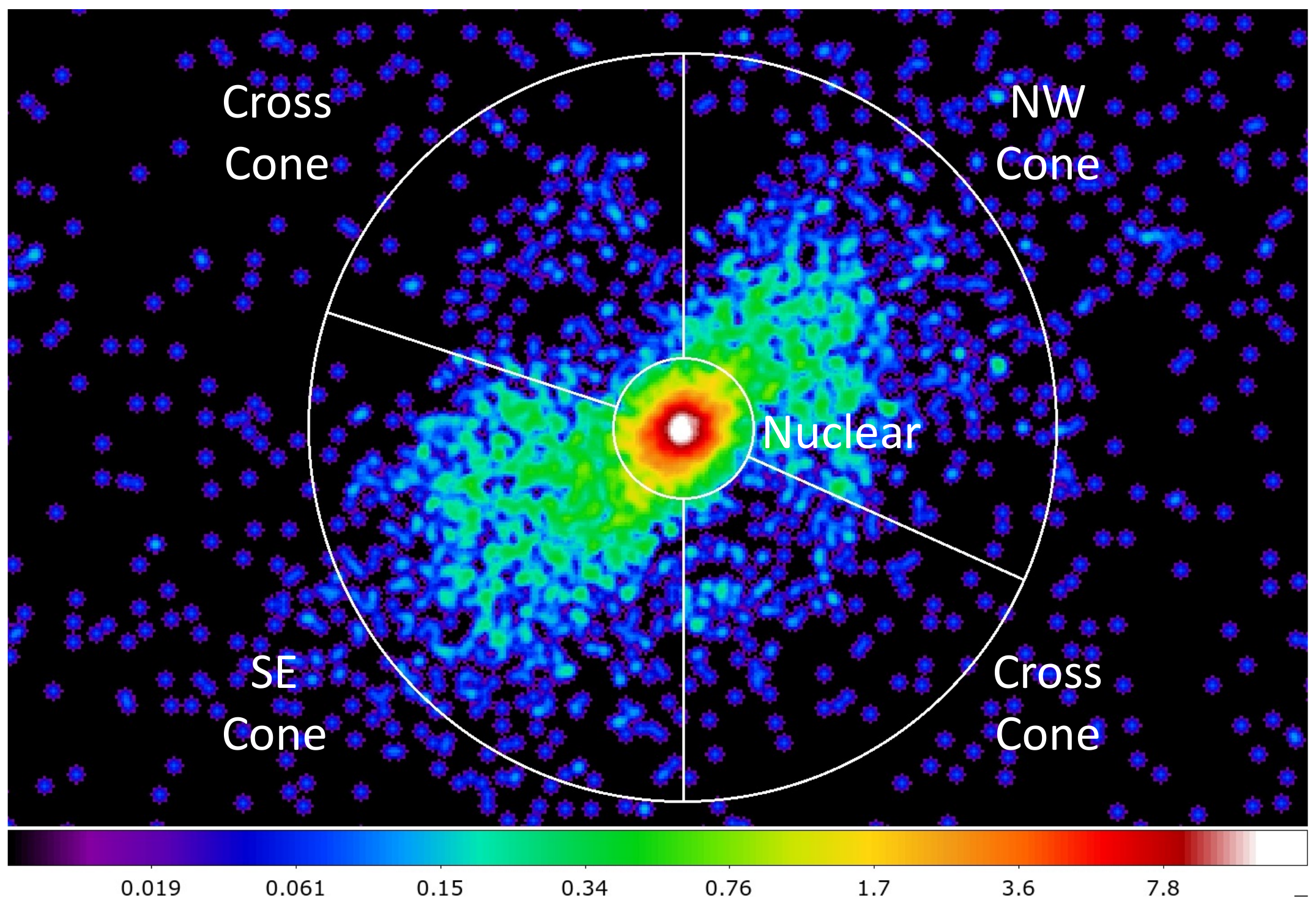}
\caption{1/8 pixel image of the emission in the 0.3–7 keV band. White solid lines divide the nucleus of NGC 5728 from the extended ($>$1.5$''$) region, which, in turn, is split into four regions to separate the bicone from the cross-cone area, as shown. Specifically, the bicone regions are enclosed in northwest (NW) and southeast (SE) sectors and limited at the position angles 336$\degree$-90$\degree$ and 162$\degree$-270$\degree$, respectively. The off-nuclear point sources have been removed for the spatial and spectral analysis, as shown.}
 \label{fig:full_band} 
 \end{figure}

Based on the cone and cross-cone profiles obtained in Section \ref{sec:images}, we selected four regions from which to extract the spectral data: (1) a 1.5$''$-radius circle to properly encompass the nuclear emission and the surrounding inner circumnuclear emission; two annuli of 1.5$''$-8$''$ radius centered on the nucleus, which do not encompass the central emission region, we name (2) northwest (NW) cone, (3) southeast (SE) cone, and two sectors of 2.5$''$-8$''$ annulus we name (4) cross-cone\footnote{Based on the radial profiles, there is no hard extended emission in the cross-cone direction ($>$ 4 keV). Therefore, to avoid any emission from nuclear spillover, we opt to use a 2.5$''$-8$''$ annulus for the cross-cone emission spectral extraction.} (Fig. \ref{fig:full_band}). Previous radio and optical studies of a different CT AGN, IC 5063, have shown that the two cone regions can host multiphase gas with different properties, e.g., density and kinematics \citep{morganti2007a,dasyra2016a}. Taking this into account, we separately analyzed the spectral properties of the NW and SE cones, to account for any differences between the two locations.\par

%The spectra extracted from the two sectors in the cross-cone direction do not show significant differences, possibly due to the smaller number of counts, and are, therefore, merged into one single spectrum, for analysis purposes.\par 

We used \textsc{ciao} \textsc{specextract} to extract the spectra from individual observations. All spectra were background subtracted from a circular region of 10$''$ radius, free of X-ray point-like sources. The eleven individual spectra and responses were co-added using the \textsc{combine\_spectra}\footnote{https://cxc.cfa.harvard.edu/ciao/ahelp/combine\_spectra.html} script and binned to have a minimum of 20 counts bin$^{-1}$, to allow for a reliable $\chi^2$ fitting. We then used \textsc{sherpa}\footnote{https://cxc.cfa.harvard.edu/sherpa/} \citep{freeman2001a} to fit the final merged source spectrum. \par

\subsection{Spectral Modeling}

Following previous works \citep[e.g.,][]{fabbiano2018a, paggi2022a}, we adopted two classes of models to describe the spectra extracted from the different regions: \par 

\bigskip

\textbf{\underline{PHENOMENOLOGICAL MODELS}}:\par 
\noindent This class of models is used to locate the spectral regions of line emission in each extracted spectra. In addition to the photo-electric absorption by the Galactic column density along the line of sight ($N_{H}$ = 7.53$\times$10$^{20}$ cm$^{-2}$, derived with the \textsc{nasa heasarc} tool\footnote{https://heasarc.gsfc.nasa.gov}), these models comprise a power-law with variable photon-index and several red-shifted Gaussian lines with widths free to vary (requiring sigma $\geq$ 0.1 keV, the ACIS-S instrument energy resolution). The redshift assumed for the emission lines is systemic only ($z$= 0.00932), and an intrinsic absorption component was included in the models when necessary. \par 

\bigskip

Table \ref{tab:parameters_pheno} (in Appendix) lists the results of the spectral fits for the phenomenological models, including the best-fit energies of the emission lines\footnote{We opt to move Table \ref{tab:parameters_pheno} to the Appendix to maintain the flow of the paper}. The total flux of the best-fit model is also shown. We note that these models should not be used to infer any "physical" processes and line fluxes, especially in the range $<$ 1.5 keV, where several blended emission lines cannot be entirely spectrally resolved at ACIS resolution.\par

\bigskip

 \textbf{\underline{PHYSICAL MODELS}:}\par
\noindent We use this class of models to probe possible physical emission mechanisms in each extraction region. The extracted spectra were fitted with photoionization (\textsc{cloudy}\footnote{http://www.nublado.org/}) \citep{ferland2017a} and thermal (\textsc{apec}\footnote{http://www.atomdb.org/}) \citep{foster2012a} models, as well as combinations of both (\textsc{cloudy+apec} models). In all the fits, the model includes the best-fit nuclear continuum and Fe K$\alpha$ line model from Table \ref{tab:parameters_pheno}. We started with one-component models and added additional components to improve the fit quality.  We also grouped the best-fit residuals to identify cases in which we obtain overall formally good fits (i.e. reduced $\chi^{2}$ $\sim$ 1) but have significant residuals in certain energy bands, to which $\chi^{2}$ is not sensitive. We used these inspections to justify the addition of a new component to the model.\par 
\medskip

\textbf{1. Photoionization}\par 
\noindent We built grids of models with the \textsc{cloudy} c17.03 package for the photo-ionization components. Assuming a continuum source with a spectral energy distribution (SED) in the form of a power law L$_{\nu}$ $\propto$ $\nu^{-\alpha}$, with $\alpha$ = 1.0, for 1$\times$10$^{-4}$ eV $<$ h$\nu$ $<$ 13.6 eV, $\alpha$ = 1.3, for 13.6 eV $<$ h$\nu$ $<$ 500 eV, and $\alpha$ = 0.5, for 500 eV $<$ h$\nu$ $<$ 30 keV, with exponential cutoffs above and below the limits \citep[e.g.,][]{kraemer2020a}. The grids of models obtained also depend on the choice of input parameters, i.e., the radial distance of the photoionized gas to the central AGN ($r$), the column density (NH), and the number density (n$_{H}$), being parameterized in terms of the ionization parameter, U.\par 

Concerning element abundances for these models, we consider 1.4x solar abundances. Specifically (in log, relative to hydrogen, by number): He=-1.00, C=-3.47, N=-3.92, O=-3.17, Ne=-3.96, Na=-5.69, Mg=-4.48, Al=-5.53, Si=-4.51, S=-4.82, Ar=-5.40, Ca=-5.64, Fe=-4.4, and Ni=-5.75 \citep[e.g.][]{trindadefalcao2021a, kraemer2020a}.\par 

The grid of \textsc{cloudy} models was built over a range of NH and U (varying in the range log U = [-2.00 : 3.00] in steps of 0.1, and the hydrogen column density NH, expressed in cm$^{-2}$, varying in the range log NH = [20.0 : 23.5] in steps of 0.1). We assumed a turbulence velocity of 100 km~s$^{-1}$ \citep[e.g.,][]{armentrout2007a, kraemer2020a}. The \textsc{cloudy} output was converted onto additive emission components (ATABLES) in a FITS format \citep{porter2006a}, with \textsc{sherpa} interpolating between the values in the grid in the fitting process.\par

\medskip

\textbf{2. Thermal}\par 
\noindent The thermal plasma was represented by \textsc{apec} models with solar abundances. The model predicts line and continuum emissivities for an optically thin, hot plasma, using the AtomDB atomic database to calculate an emission spectrum from a collisionally-ionized diffuse gas. The modeled optically thin emission can result from thermalized ISM after being collisionally ionized via interactions with winds or radio jets from either the nuclear source or the very prominent star-forming ring. \par

\medskip

\textbf{3. Photoionization + Thermal Models}\par 
\noindent To investigate models that would give a better fit over the entire spectral range, we tried a set of mixed photoionization and thermal models (in addition to the nuclear continuum model). These included an increasing number of photoionization components (from 1 to 3) plus a thermal component and an increasing number of thermal components (from 1 to 3) plus a photoionization component. We also try a 2-photoionization + 2-thermal model. \par

\begin{table}[htb!]
\footnotesize
\begin{center}
\caption{\textbf{Best Physical Parameters of the Photo-ionized and/or Collisionally-ionized Gas in All Regions}}
\label{tab:parameters_models} 

\begin{tabular}{lccccccc}
\multicolumn{8}{c}{}\\
\hline
\multicolumn{1}{l}{Region}
&\multicolumn{1}{c}{Best-fit Model}
&\multicolumn{1}{c}{Model Energy Flux$^{*}$}
&\multicolumn{1}{c}{Reduced $\chi^{2}$}
&\multicolumn{1}{c}{d.o.f.}
&\multicolumn{1}{c}{Alternative Model Components}
&\multicolumn{1}{c}{Reduced $\chi^{2}$}
&\multicolumn{1}{c}{d.o.f.}
\\
&  &(0.3-7.0 keV; in erg/cm$^{2}$/s) & & & \\
\hline

& log(NH1)=20.4$\pm$0.1& & & & log(NH1)=20.3$\pm$0.1 & \\
\textbf{(1.5$''$ circle)} & log(NH2)=23.3$\pm$0.5 & 6.8$\times$10$^{-13}$ &0.68&213&log(NH2)=23.3$\pm$0.4 &0.67&211\\
& log(U1)=0.4$\pm$0.1& & & &log(U1)=0.4$\pm$0.1  &\\
& log(U2)=-1.6$\pm$0.1& & & &log(U2)=-1.6$\pm$0.1 &\\
& & & & &$kT$1=1.4$\pm$0.2  &\\
& & & & &EM1=2.1$\times$10$^{-6}$ &\\

\hline

& log(NH1)=21.1$\pm$0.1 &  & &  \\
 & log(NH2)=20.5$\pm$0.1&   & &\\
& log(U1)=1.7$\pm$0.1& &   & & log(NH1)=21.2$\pm$0.12 &\\
\textbf{(NW Bicone Region)}& log(U2)=0.4$\pm$0.1 &6.7$\times$10$^{-14}$ &0.88&41&log(NH2)=20.6$\pm$0.12 &  0.93&45\\
&$kT$1=0.6$\pm$0.15& & & & log(U1)=1.6$\pm$0.1  &\\
&$kT$2=1.4$\pm$0.27& & & & log(U2)=0.3$\pm$0.10 &\\
& EM1=1.5$\times$10$^{-6}$& & & &\\
& EM2=1.8$\times$10$^{-6}$& & & &\\

\hline

&  & & & &log(NH1)=21.7$\pm$0.1 &  \\
& log(NH1)=21.7$\pm$0.1 & & & &log(NH2)=21.7$\pm$0.1 &\\
\textbf{(SE Bicone Region)} &log(U1)=-0.9$\pm$0.08 & 1.0$\times$10$^{-13}$&0.94&60& log(U1)=-0.9$\pm$0.10 &  0.98&57\\
&$kT$1=0.8$\pm$0.11& & && log(U2)=0.5$\pm$0.11 &\\
&$kT$2=1.3$\pm$0.11& & && $kT$1=0.8$\pm$0.12 &\\
&EM1=5.3$\times$10$^{-6}$ & & && $kT$2=1.4$\pm$0.12 &\\
&EM2=1.6$\times$10$^{-6}$ & & && EM1=5.4$\times$10$^{-6}$ &\\
& & & && EM2=1.6$\times$10$^{-6}$ &\\

\hline
 \hline
\end{tabular}
\end{center}
U is the ionization parameter of each component\\
NH is the column density (cm$^{-2}$) of each component\\
kT is the temperature (keV)\\
EM is the normalization of the \textsc{apec} model, which is proportional to the emission measure.\\
$^{*}$ Calculated using the \textsc{calc\_energy\_flux} tool on \textsc{sherpa}. It represents the integral of the unconvolved model over the 0.3-7 keV energy band.
\end{table}

\section{Results}
\label{sec:results}

\subsection{X-ray Morphology}

As shown in Figure \ref{fig:radial_1}, the soft 0.3-1.5 keV band is the most extended. Comparison with the PSF profiles shows that the surface brightness in this energy band is extended in both cone and cross-cone directions, although the extent in the cross-cone direction is smaller. In the cone direction, the surface brightness can be traced to $\sim$ 10$''$ ($\sim$ 2 kpc) radially and up to $\sim$ 7$''$ (1.4 kpc) in the cross-cone direction. In the 1.5-3.0 keV band, emission can be traced to $\sim$ 10$''$ (2 kpc) radially, in the cone direction, and up to $\sim$ 6$''$ (1.2 kpc) in the cross-cone, in excess of the respective model PSF profiles.\par

 Fig. \ref{fig:3-6kev} (top) shows the cone and cross-cone profiles for the 3-6 keV band, which contains 567$\pm$24 net source counts within a 1.5$''$-8$''$ radius annulus, after excluding the off-nuclear point sources in the field. There are $\sim$ 462 counts in excess of what could originate from a nuclear point source (Table \ref{tab:excess}). The surface brightness in this energy band extends radially to $\sim$ 10$''$ (2 kpc) in the cone direction; therefore, this continuum emission cannot be directly attributable to the emission from the immediate vicinities of the central active nucleus. The extent of the surface brightness decreases at higher energies and in the cross-cone direction. In all cases, the emission in the cone direction is more extended than that of the PSF in the same energy band, but it is consistent with the PSF for energies 3-6 keV in the cross-cone direction. \par 
 
 In \citealt{fabbiano2017a}, the authors reported the discovery of kpc-scale diffuse emission in both the hard continuum (3-6 keV) \textit{and} in the Fe K$\alpha$ line band in ESO 428-G014, another CT AGN. Fig. \ref{fig:3-6kev} (bottom) shows the imaging cone and cross-cone X-ray emission and surface brightness radial profiles in the 6.1-6.6 keV band for NGC 5728. As shown, this emission is extended in the cone direction, $\sim$ 6$''$ (1.2 kpc), while it is consistent with the PSF in the cross-cone.  \par 

In summary, radial profiles and images reveal kpc-scale emission at energies $<$3 keV, which are dominated in X-rays by line emission (see Section \ref{sec:spec_analysis}), and in the 3-6 keV continuum and Fe K$\alpha$ line energy bands. While this extent is more pronounced in the cone direction, which is the direction of the ionization bicone, the emission is also extended in the "cross-cone" direction at energies up to $\sim$ 3 keV. \par

\subsection{Spectral Properties}
We characterized the spectral properties of the extended and nuclear X-ray emission in NGC 5728. The results of our detailed analysis are shown below for each extraction region. \par

\subsubsection{The Nuclear Emission}
\label{sec:nuc_emission}

The nuclear spectrum (1.5$''$ circle) exhibits both strong hard ($>$3 keV) and soft X-ray emission at lower energies. We fit the hard emission by employing a nuclear model that includes three components: (i) one component to account for the SMBH emission transmitted through the dusty torus, (ii) one component to account for the reflected SMBH emission (due to Compton scattering) from the torus, and (iii) one model component to account for the 6.4 keV Fe K$\alpha$ emission line. In addition, we employed additional spectral components to model the soft excess, as described below \citep[e.g.,][]{krolik1994a}.\par 

We used \textsc{sherpa} to describe the AGN spectrum using the following set of models. The emission from the SMBH was assumed to follow a power law with a high-energy cutoff, modeled with \textsc{xszcutoffpl}. The cutoff energy in AGNs is usually above 150 keV, but for this study we fixed it at 200 keV \citep{semena2019a}. The absorption of the emission in the torus was modeled using the \textsc{xsztbabs} model. The reflected emission was described using the \textsc{xspexrav} spectral model, which is a semi-analytical model of the spectrum produced when an isotropic X-ray emission is reflected from a plane layer of cold, non-ionized gas with infinite optical depth \citep{magdziarz1995a}. The observer's angle to the plane layer was assumed to be cos$\theta$= 0.5, and the iron abundance in the layer was set to be solar. The cutoff energy in the incident emission spectrum E$_{\rm cut}$ was linked to that of the reflected emission. To model the fluorescent iron line (Fe K$\alpha$), we used one gaussian with fixed line position at 6.41 keV, given the AGN redshift. In \textsc{sherpa} terminology, our model for the hard spectrum is expressed as \textsc{xstbabs} $\times$ [\textsc{xspexrav} + \textsc{xsztbabs} $\times$ [\textsc{xszcutoffpl}] + \textsc{xszgauss}]. As for the soft emission, we fit the observed soft excess with a simple absorbed power law (\textsc{xspowerlaw}).\par 

The best-fit model yields a soft power-law component, a hard power-law component, and a reflection component with photon indexes $\Gamma$ = 1.5 $\pm$ 0.2, $\Gamma$ = 1.8 $\pm$ 0.1, and $\Gamma$ = 1.8 $\pm$ 0.1, respectively. The estimated column density absorption is N$_{H}$ $\sim$ 1.0 $\pm$ 0.1 $\times$ 10$^{24}$ cm$^{-2}$, which is consistent with the findings of \citet{shu2007a} and \citet{semena2019a} for this galaxy. \par 

There are a total of nine fitted emission lines, including neutral Fe K$\alpha$ emission with an equivalent width EW=0.77 keV, similar to that found in previous works \citep[e.g.,][]{semena2019a, tanimoto2022a}. Table \ref{tab:parameters_pheno} gives the best-fit parameters and the energies of the fitted emission lines. \par 

 The nuclear spectrum shows emission to the left ($\sim$ 6.1 keV) and to the right ($\sim$ 6.7 keV) of the neutral Fe K$\alpha$ line, which our phenomenological fit did not pick up, as well as a Fe edge absorption feature at 7.1 keV. Below 3 keV, the most prominent feature is at $\sim$ 1.8 keV, which we associate with a blend of Mg~XII, Si~XIII, and Fe~XXIV transitions. For comparison, we fit the hard ($>$ 3 kev) nuclear spectrum with a second physically motivated model, \textsc{borus02} \citep{balokovic2018a}, which also incorporates hard ($>$ 8 keV) constraints. Our results show that the excess emission observed to the left and to the right of the neutral Fe K$\alpha$ line are roughly identical, compared to the results obtained using \textsc{pexrav}, suggesting these features to be model-independent. We present the details of this analysis in the Appendix (Fig. \ref{fig:borus}). \par

\begin{figure}[htb!]
  \centering
\begin{minipage}[b]{\textwidth}
  \includegraphics[width=18cm]{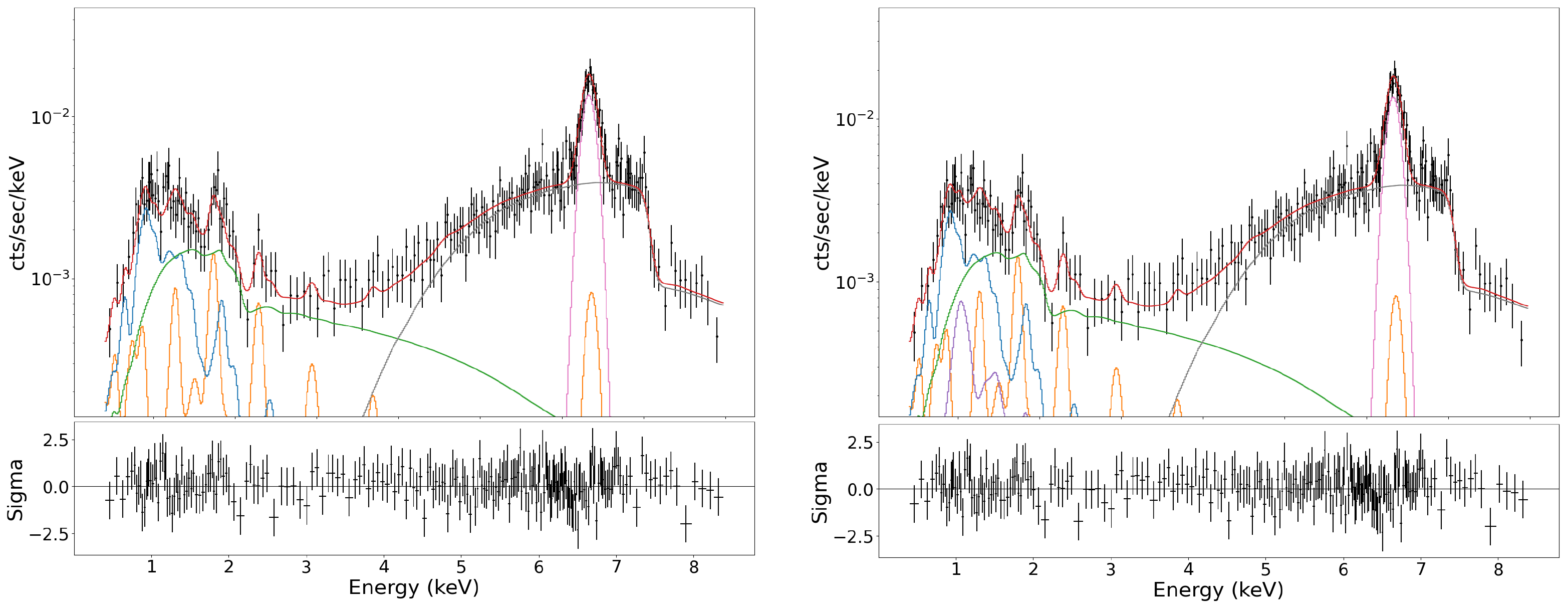}
 \end{minipage}\qquad
\caption{Spectrum from the inner 1.5$''$-radius circle with the best-fit (top) and best-fit residuals (bottom) for a 2-photoionization component model (left) and a 2-photoionization+1-thermal models (right). The hard X-ray spectral fit consists of a Gaussian emission line to model the Fe–K$\alpha$ transition (pink line), and an intrinsically absorbed reflection  component \textsc{pexrav} with $\Gamma$ = 1.8 + a cutoff power law (modeling the incident emission) with $\Gamma$ = 1.8 (gray line). The soft X-ray emission is best fit with an absorbed power-law component with $\Gamma$ = 1.5 (green line), plus 2- photoionization components (blue and orange lines). Adding a third thermal component (purple line, right panel) does not improve the quality of the fit.}
 \label{fig:nuc_spec} 
 \end{figure}

Multi-component models combining photoionization and/or optically thin thermal models show that the soft nuclear emission is dominated by a high photoionization (log U $\sim$ 0.4), low column density (log NH $\sim$ 20.3 cm$^{-2}$) gas component, which allows us to reproduce prominent emission lines at $\sim$1.8 keV and $\sim$2.3 keV. A second, low-photoionization component (log U $\sim$ -1.6), with high column density (log NH $\sim$ 23.3 cm$^{-2}$) models the soft excess emission (Fig. \ref{fig:nuc_spec}). Best fit parameters for chosen models are given in Table \ref{tab:parameters_models}.\par 

We do not find any evidence that suggests the need for a thermal component to fit the nuclear spectrum. In fact, as shown in Fig. \ref{fig:nuc_spec}, the addition of a 1-thermal component to a 2-photoionization model does not modify the residuals or reduced $\chi^{2}$ of the fit (Table \ref{tab:parameters_models}).  \par 

\subsubsection{The Ionized Bicone}
\label{sec:BD_emission}

Fig. \ref{fig:merged} and \ref{fig:radial_1} clearly show the presence of X-ray emission stretching SE–NW along the bicone direction out to $\sim$ 2 kpc (10$''$) from the nucleus. This emission is particularly prominent in the soft band. Extended soft X-ray emission has been observed in numerous Seyfert 1.5–2 galaxies, spatially correlated with [O~III] $\lambda$5007 emission \citep[e.g.,][]{levenson2006a, fabbiano2022a}, suggesting a common origin primarily due to photoionization, and partially to collisional ionization of circumnuclear clouds \citep[e.g.,][]{wang2011b}. Prominent emission lines below 3 keV, typically observed in CT Seyferts \citep{wang2011c, fabbiano2017a, maksym2019a}, are also present in the analyzed bicone spectra, including Ne~IX, Mg~XII, and Si~XIII. Given the ACIS spectral resolution, these lines are typically blended in the observed spectra. \par

Fig. \ref{fig:radial_2} and \ref{fig:3-6kev} show that the bicone emission is also present at energies $>$3 keV. Harder extended components (in both the continuum emission above 3 keV and the neutral 6.4 keV Fe K$\alpha$ line) have been detected with \textit{Chandra} in AGNs \citep[e.g.,][]{fabbiano2017a}. Table \ref{tab:excess} gives the excess counts over the PSF in each conical sector for different energy bands. \par

For each spectrum, we fit the continuum with a power-law component ($\Gamma$ = 1.5 for both regions, with intrinsic absorption NH=4.4$\times$ 10$^{20}$ cm$^{-2}$ for the NW cone). The prominent emission lines can be fitted with the addition of redshifted ($z$=0.00932) Gaussians (nine to the NW cone spectrum, ten to the SE cone spectrum), as listed in Table \ref{tab:parameters_pheno}. The phenomenological fit returns $\chi^{2}$ $\sim$ 0.27, and 0.85, for the NW and SE bicone, respectively. Significant residuals can be seen to the left and right of the 6.4 Fe K$\alpha$ line. \par 

As before, we also examined the physical mechanisms responsible for the X-ray emission in the extended conical regions. We initially fit the spectra with a single physical component, adding additional components as necessary. The NW cone is best fit with a 2- photoionization + 2- thermal model. A 2- photoionization component model also returns a good fit but fails to fit the emission $\sim$ 1.0 keV. All models (single and/or multi-component models) fail to fit the emission detected at $\sim$ 6 keV to the red of the neutral Fe K$\alpha$ line. \par 

\begin{figure}[htb!]
  \centering
\begin{minipage}[b]{\textwidth}
  \includegraphics[width=18cm]{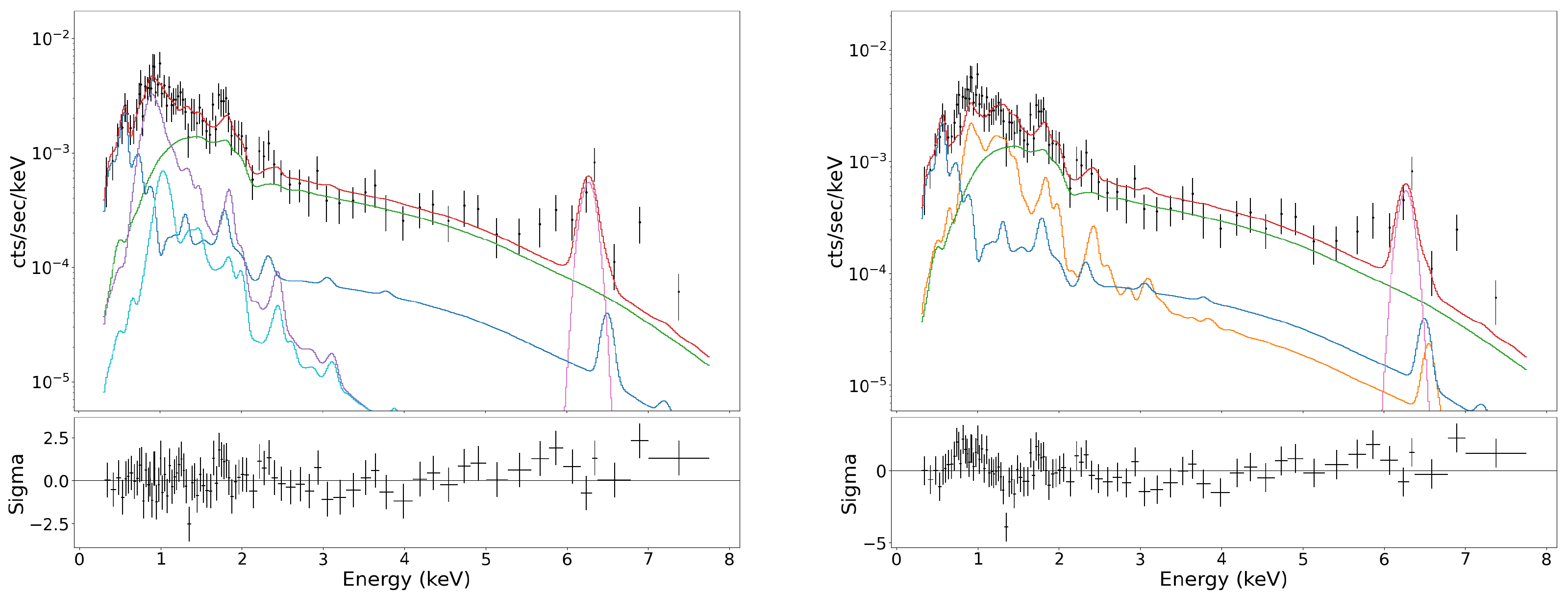}
 \end{minipage}\qquad

\caption{Spectrum from the SE cone region with best-fit (top) and best-fit residuals (bottom). The soft ($<$3 keV) X-ray emission is best fit with an absorbed power-law component with $\Gamma$ = 1.5 (green line), plus a 1- photoionization + 2-thermal models (blue, purple, and cyan lines, respectively; left panel). The hard ($>$3 keV) X-ray spectral fit consists of a Gaussian emission line to model the Fe–K$\alpha$ transition (pink line). Even though the soft emission can also be fitted with 2- photoionization model (right panel), the addition of thermal components to the 1- photoionization model reduces the residuals $<$ 2 keV and returns the best reduced $\chi^{2}$.}
 \label{fig:BD_S_spec} 
 \end{figure}

The SE cone exhibits similar features to the NW cone spectrum, except for a less intense emission peak at $\sim$ 2 keV and a less prominent Fe K$\alpha$ emission line. Fitting all the emission lines requires at least 1- photoionization component + 1- thermal component. Using photoionization \textit{or} thermal models only fails to fit emission $\sim$ 1.3 keV and below 1 keV, as shown in Fig. \ref{fig:BD_S_spec}.\par

\begin{figure}[htb!]
  \centering

 \includegraphics[width=10cm]{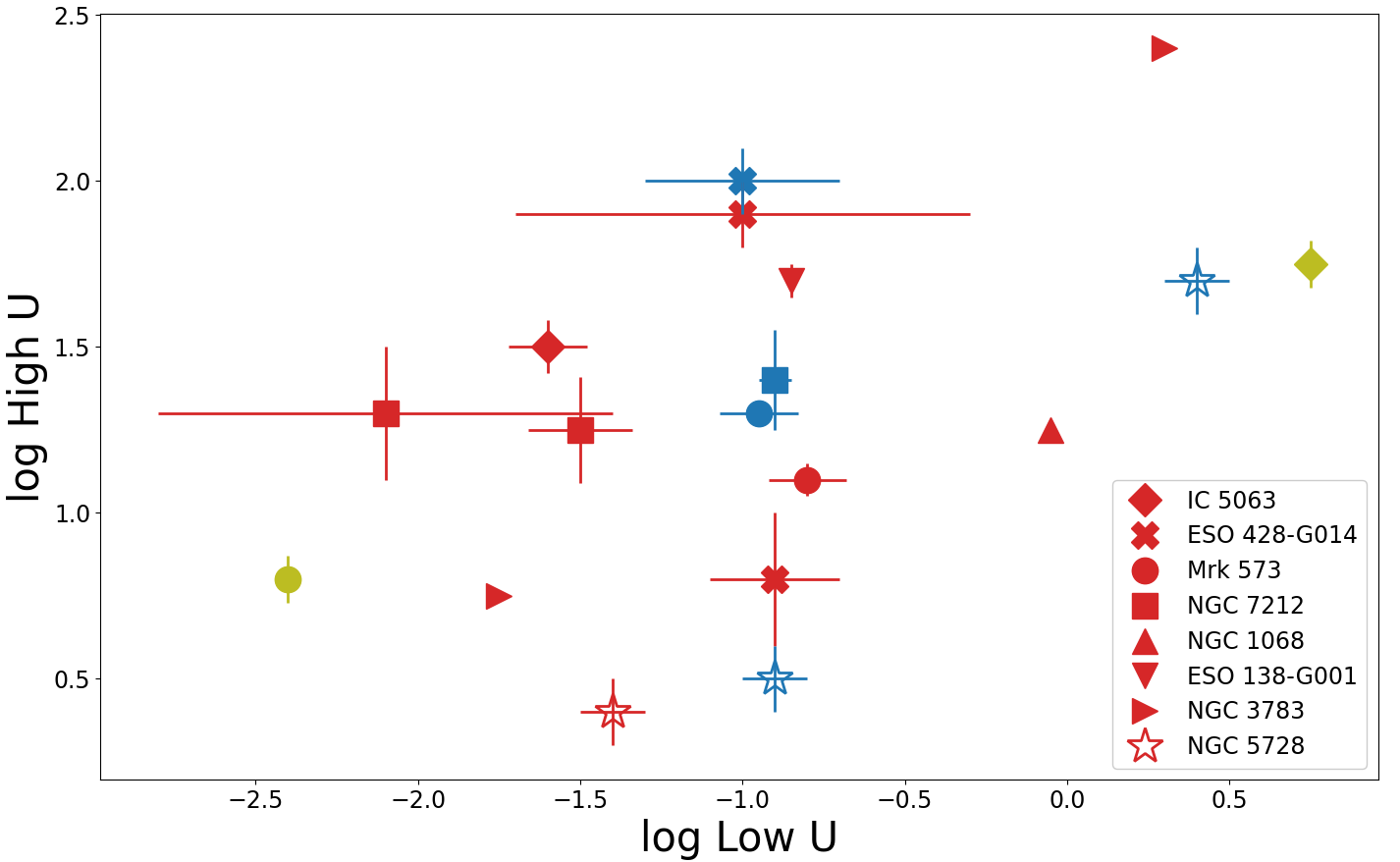}
\caption{Comparison between the low (x-axis) and high (y-axis) ionization parameters for the best-fit models for NGC 5728 (stars) and the best-fit parameters derived for other Seyfert galaxies from the literature. The icons in red represent the values for the ionization parameters for the nuclear region, blue icons represent the values in the bicone, and green icons the values in the cross-cone direction. The numbers for the best-fit parameters for the Seyfert galaxies in the plot were obtained from \citealt{travascio2021a} (IC 5083), \citealt{fabbiano2018a} (ESO 428-G014), \citealt{paggi2012a} (Mrk 573), \citealt{jones2020a} (NGC 7212), \citealt{kraemer2015a} (NGC 1068), \citealt{decicco2015a} (ESO-138-G001), and \citealt{blustin2002a} (NGC 3783).}
 \label{fig:ion_param} 
 \end{figure}

In Fig. \ref{fig:ion_param}, we compare the low and high ionization parameters (U) used to model the spectra from different Seyfert galaxies, including the values from this work. The ionization parameters we obtain for NGC 5728 are compatible with those reported in the literature for the nuclear and bicone regions \citep[e.g.,][]{paggi2012a, fabbiano2018a, jones2021a}.

 \begin{figure}[htb!]
  \centering
\begin{minipage}[b]{\textwidth}

  \includegraphics[width=18cm]{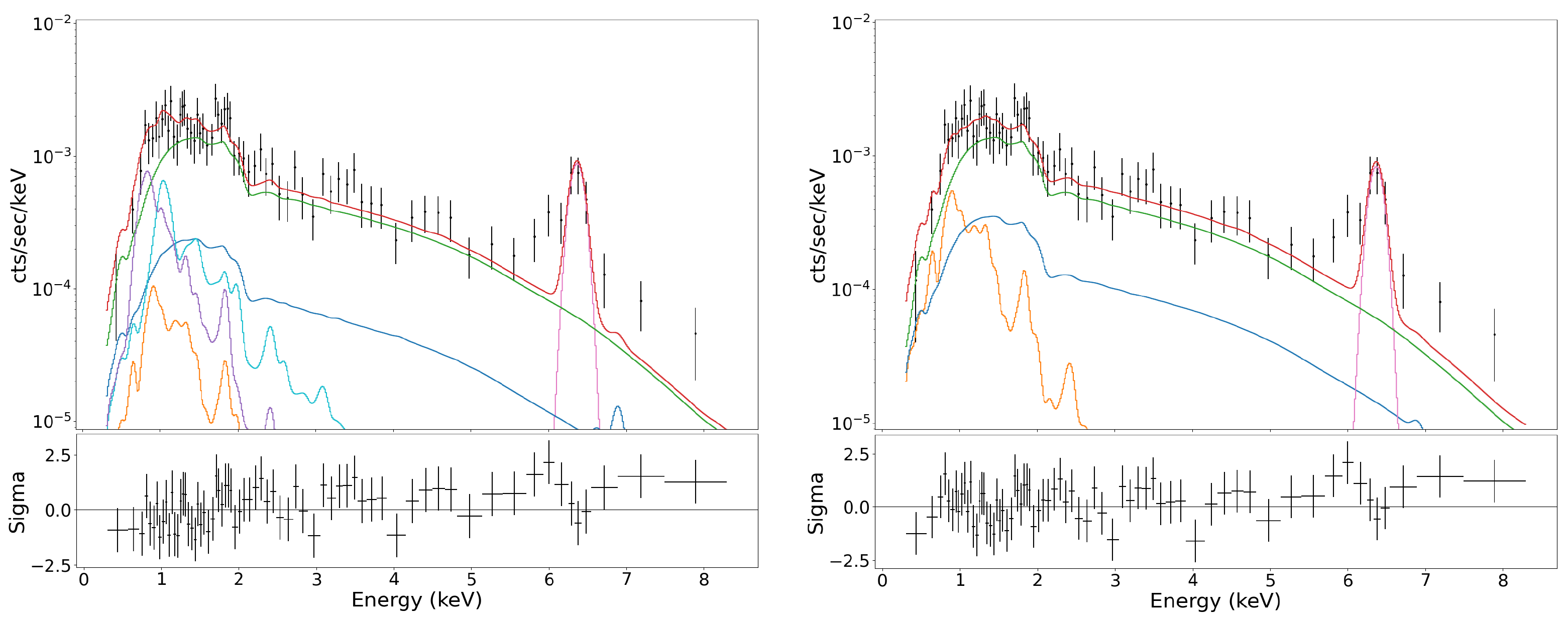}
 \end{minipage}\qquad
\caption{Spectrum from the NW cone region with best-fit (top), and best-fit residuals (bottom). The soft ($<$3 keV) X-ray emission is best fit with an absorbed power-law component with $\Gamma$ = 1.5 (green line), plus a 2- photoionization + 2-thermal models (blue and orange lines, purple and cyan lines, respectively; left panel). The hard ($>$3 keV) X-ray spectral fit consists of a Gaussian emission line to model the Fe–K$\alpha$ transition (pink line). Even though the soft emission can also be fitted with 2- photoionization model (right panel), the addition of 2- thermal components reduces the residuals $<$ 2 keV.}
 \label{fig:BD_N_spec} 
 \end{figure} 
 
\subsubsection{The Cross Cone Emission}
\label{sec:cross_cone}
Fig. \ref{fig:radial_1} and Table \ref{tab:excess} show significant extended emission in the cross-cone region in the soft band. This emission is evident at energies $<$3 keV, extending out to a radius of $\sim$ 1.4 kpc (7$''$) from the nucleus, but it is not seen above 4 keV. Therefore, to avoid contamination from the PSF's wings, we analyze the cross-cone emission only in the 0.3-3 keV energy band.\par 

%Extended soft X-ray emission, perpendicular to the bicone axis, has been observed in other AGN \citep[e.g.,][]{wang2011c, fabbiano2017a, maksym2019a, fabbiano2022a}. This emission is not expected from the classical AGN unification paradigm \citep{antonucci1993a}. A possible explanation relies on the fact that the nuclear torus may be porous, allowing part of the nuclear photoionizing continuum to escape \citep{kraemer2008a, nenkova2008a, fabbiano2018a} (see Section \ref{sec:contribution}). \par 

 The best phenomenological model for the cross-cone emission consists of an absorbed (intrinsic absorption, N$_{H}$= 4.2$\times$10$^{21}$ cm$^{-2}$), steep ($\Gamma$=1.3) power law plus four redshifted ($z$=0.00932) Gaussians. The spectrum still exhibits the strong emission at $\sim$1.8 keV observed both in the nuclear and bicone spectra and at $\sim$ 0.9 keV. \par 

\begin{figure}[htb!]
  \centering
\begin{minipage}[b]{\textwidth}
%\hspace*{-1.25cm} 
  \includegraphics[width=18cm]{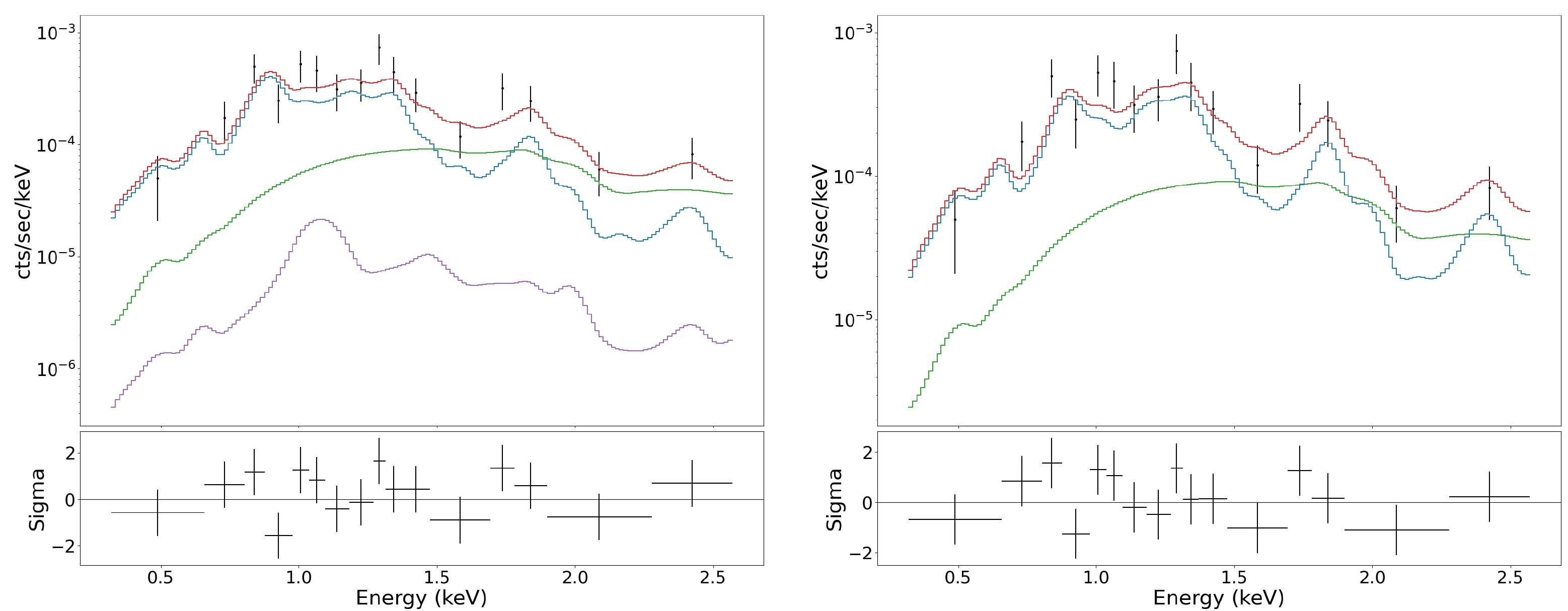}
 \end{minipage}\qquad

\caption{Spectrum for the cross-cone region with best-fit (top) and best-fit residuals (bottom). The soft ($<$3 keV) X-ray emission is best fit with an absorbed power-law component with $\Gamma$ = 1.3 (green line), plus a 1- 
 photoionization + 1- thermal model (blue and purple lines, respectively; left panel). Even though the soft emission can also be fitted with different single-component models (right panel), adding a thermal and/or photoionization component improves the reduced $\chi^{2}$.}
 \label{fig:CCD_spec} 
 \end{figure}

The cross-cone spectrum can be best fit with at least 1- photoionization + 1- thermal components. The photoionization component has log U $\sim$ 0.3 and column density log (NH)= 23.3 cm$^{-2}$. Fig. \ref{fig:CCD_spec} shows that single-component alternative models, such as 1- photoionization (Fig. \ref{fig:CCD_spec}, right panel), fail to fit the emission at $\sim$0.7 keV. We opt not to show the results of the best-fit models for the cross-cone emission in Tables \ref{tab:parameters_models} and \ref{tab:parameters_pheno} since the number of counts present in the spectrum is not enough to obtain a reliable fit with good statistics.  \par 

Fig. \ref{fig:radial_1} shows that the radial profile in the cross-cone direction presents two "bumps" (one in the NE cross-cone and the other in the SW cross-cone) in the 0.3-1.5 keV band. These features can be easily identified in the 0.3-1.5 keV image as arch-like structures at $\sim$ 1 kpc from the nucleus, consistent with the location of the star-forming ring detected by \citet{durre2019a} in this object. We will explore these features more fully in a subsequent paper.\par 

\subsubsection{Unidentified Spectral Features}
\label{sec:red_blue_wings}

As reported in Table \ref{tab:parameters_pheno}, we find three unusual emission features in AGN X-ray spectra. The first feature is characterized by emission at $\sim$ 2.8 keV, which is observed in both cone emission spectra. As in \citet{jones2020a}, we suggest the identification of this line as being an argon fluorescence line (E$_{\rm lab}$= 2.96 keV). \par 

We also detect emission features at $\sim$ 3.5 keV in all regions and at 4.5 keV in the NW and SE cone regions. Possible identifications for these lines include Ca K$\alpha$ fluorescence line (E$_{\rm lab}$=3.69), and Ar~XVII (E$_{\rm lab}$=3.69). These lines have been previously observed by \citet{jones2020a} in NGC 7212, another nearby ($z$=0.0266) CT AGN.\par

We find two distinct emission lines near 6 and 7 keV, which are observed in the nuclear, NW cone, and SE cone spectra. These features appear as residual excess "wings" redward and blueward of the neutral Fe~K$\alpha$ emission line in Fig. \ref{fig:nuc_spec}, \ref{fig:BD_N_spec}, and \ref{fig:BD_S_spec}. In the physically motivated models, the "blue wing" is partially modeled by high photoionization components as Fe~XXV 6.7 keV emission, but with significant fit residuals, as shown on the bottom panels of Fig. \ref{fig:BD_N_spec}, and \ref{fig:BD_S_spec}. On the other hand, the "red wing" feature cannot be modeled by any combination of photoionization and/or thermal models and appears as excesses in all fits. Similar features have been recently reported in \textit{Chandra} ACIS imaging spectroscopy of the nucleus of the Seyfert 2 Galaxy
Mrk 34 \citep{maksym2023a}. In \citealt{trindadefalcao2023b}, we conducted a detailed analysis of this unmodeled emission in the bicone. Based on the count excess over the best-fit model, we estimate that, in this region, the feature to the [left, right] of the neutral Fe K$\alpha$ line has [6.5$\sigma$, 5.3$\sigma$] significance, with an equivalent width = [1.76 keV, 2.60 keV]. A likelihood ratio test based on a simulation with 100,000 iterations yields a probability p$<$1$\times$10$^{-5}$ that a simple power-law continuum plus narrow Fe K$\alpha$ line is a better representation of the data than a model that includes the wings emission.\par 

%This "red wing" has been observed in other X-ray spectra of AGN \citep[e.g.,][]{turner2002a, reeves2005a, turner2008a, jones2020a}, but has not received a successful and definitive identification. One possibility is that this emission is due to redshifted and blueshifted Fe~K$\alpha$ fluorescence, in the form of AGN outflow. Some of this emission could be associated with a hot collisionally ionized plasma, which also produces highly ionized Fe~XXV, as shown in Table \ref{tab:parameters_pheno}. Another possibility is that the red wing feature is simply scattered emission-line photons, forming what is called a ‘Compton shoulder’ \citep{yaqoob2010a}. \par 

%Figure \ref{fig:red_blue} shows a comparison between the emission in the energy bands of these "red" (right) and "blue" (left) wings, for the merged image. The emission in the 5.8-6.2 keV band is extended in the cone direction, as confirmed by the excess counts over the PSF shown in Table \ref{tab:excess}. The same is true for the emission in the 6.7-7.1 keV band, however, with a smaller excess counts over the PSF. These features will be analyzed in detail in a  subsequent paper. 

% \begin{figure}
%  \centering
%\begin{minipage}[b]{.4\textwidth}
%\hspace*{-1.25cm} 
%  \includegraphics[width=9cm]{58_62_comp.pdf}
% \end{minipage}\qquad
% \begin{minipage}[b]{.4\textwidth}
%  \includegraphics[width=9cm]{67_71_comp.pdf}
% \end{minipage}\qquad \\
%\caption{Emission in the 5.8-6.2 keV band (right), and in the 6.7-7.1 keV band (left). The images are in log intensity scale. }
% \label{fig:red_blue} 
% \end{figure} 
 
\section{Discussion}
\label{sec:discussion}
Recent studies on obscured AGNs \citep[e.g.,][]{fabbiano2017a, jones2020a, travascio2021a} have produced important and surprising results, revealing extended X-ray emission in both soft X-rays and in the hard energy band, i.e., $>$ 3 keV. These have also enforced the importance of deep ACIS-S \textit{Chandra} observations, with high-resolution (sub-pixel), to obtain an overall picture of the AGN-galaxy interaction and its effects in the ISM. \par 

To further our understanding of the relationship between AGNs and their host galaxy, we conducted a detailed analysis of the soft and hard X-ray emissions in NGC 5728. Below, we discuss the implications of our imaging and spectral analysis. 

\subsection{Extent as a Function of Energy}
\label{sec:extent}

In \citet{fabbiano2017a}, the authors reported that the extent of the X-ray emission in ESO 428-G014 increases with decreasing photon energy. They suggested that this phenomenon might be caused by a larger concentration of dense molecular clouds in the central region since these clouds are responsible for scattering and reflecting the high-energy photons from the nucleus in the galaxy disk. The same effect was confirmed in more recent studies by \citet{jones2020a, travascio2021a, jones2021a}.\par 

To measure the full extent of the diffuse emission, we followed the approach adopted in \citealt{fabbiano2017a} to measure the width at which the background-subtracted surface brightness (from the radial profiles shown in Fig. \ref{fig:radial_1}, \ref{fig:radial_2}, and \ref{fig:3-6kev}) is consistent with the background surface brightness in the same energy band. This corresponds to 1\% surface brightness of the peak emission in each energy bin, and errors are primarily due to the uncertainty derived from the bin sizes. This method has been validated by several works \citep[e.g.,][]{fabbiano2018a, jones2020a, travascio2021a, jones2021a}, proving itself to be a reliable comparison of the effective emission extent as a function of energy since the radial profiles have good statistics significance at this surface brightness percentage. \par 

\begin{figure}[htb!]
  \centering
\begin{minipage}[b]{\textwidth}
%\hspace*{-1.25cm} 
  \includegraphics[width=18cm]{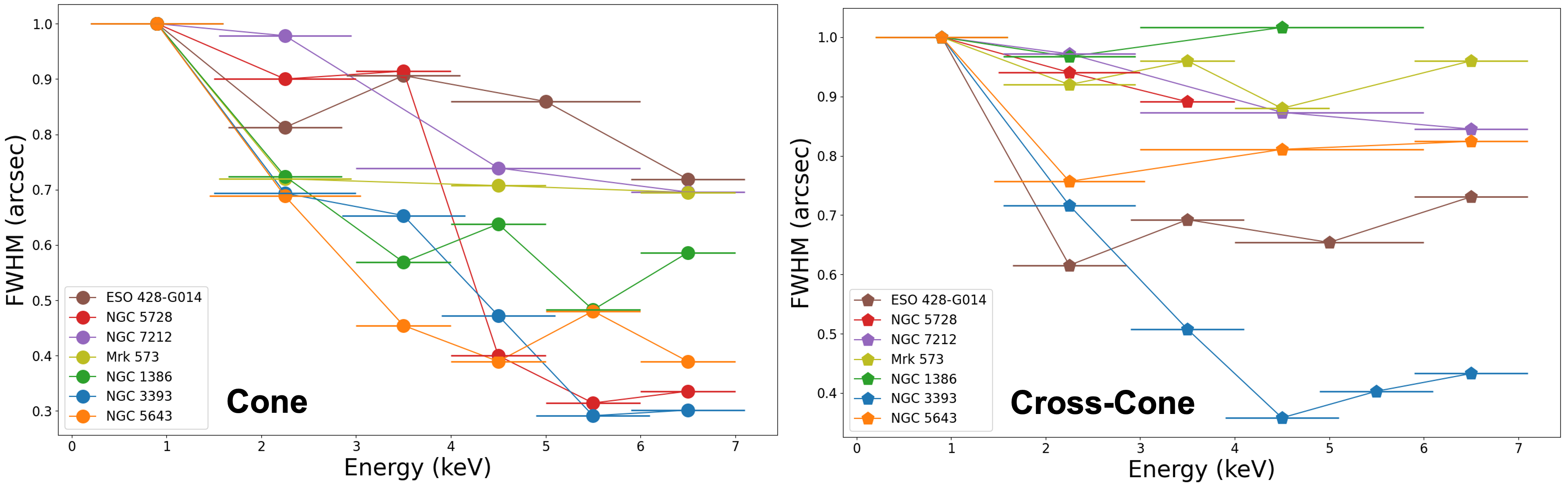}
 \end{minipage}\qquad
\caption{Comparison between the normalized extent in arcsecond at 1\% of the surface brightness for NGC 5728 and other CT AGN as a function of energy. The left panel shows the cone direction and the right panel shows the normalized FWHM in the cross-cone direction. Adapted from \citealt{fabbiano2018a} and \citealt{jones2021a}.}
 \label{fig:FWHM} 
 \end{figure}

Fig. \ref{fig:FWHM} shows a comparison between the large-scale extent of the emission in NGC 5728 in different energy bands, with that of other CT AGN, reported in the literature \citep{fabbiano2018a, jones2021a}. To facilitate the comparison, we normalized the extent of the emission in these sources to that in the soft band and used a simple line model to get the slopes of each extent-energy relation. The energy profile for NGC 5728 shows an overall trend of decreasing FWHM with increasing energy in the cone direction, as for the other analyzed AGN, but also in the cross-cone direction, unlike most other targets plotted. In the cross-cone direction (right panel), there is a decrease of the FWHM, going from 10.1$''$, at the lowest energies, to 9.0$''$ at $\sim$ 3.5 keV (before normalization). Because the emission in this work is consistent with the PSF in the cross-cone direction for energies between 4-7 keV, we opt not to plot these values. In the cone direction (left panel), the FWHM decreases from 14$''$ at the lowest energies to 4.7$''$ at $\sim$ 7 keV (before normalization). \par 

The dependence of the emission extent with energy observed in Fig. \ref{fig:FWHM} can be related to the abundance of high-density ISM clouds near the AGN and the orientation of the ionization bicone with respect to the host galaxy, as the bicone itself originates from the interaction of ionizing photons with the ISM in the host galaxy disk. Soft X-rays are thought to originate as a consequence of photoionization of the ISM and/or collisional ionization when a jet is present \citep[e.g.,][]{paggi2012a}. In contrast, the origin of hard X-rays lies in the interaction between dense ISM clouds and photons from the AGN \citep[e.g.,][]{reynolds1997a}. Our findings show that the soft emission in NGC 5728, as in the other AGN considered, extends farther than the hard emission in the cone direction (Fig. \ref{fig:FWHM}, left panel), which suggests a radial dependence on the density and size of the ISM molecular clouds near the AGN, such as a higher prevalence of optically thicker clouds closer to the nucleus \citep{fabbiano2018a}. \par 

If the ionization bicone is completely aligned with the disk of the galaxy, one would expect to see a radial dependence of the emission extent with energy in the cone direction, as observed in the left panel, but no dependence in the cross-cone direction, since the photons from the AGN do not interact with the ISM clouds at larger radii (in the Unified Model picture). Similarly, if the ionization bicone is perpendicular to the galaxy disk, one would expect no dependence in the cone direction. Even though it is possible for the cross-cone to exhibit some radial energy dependence in this scenario, this emission would be attenuated by the torus. In the case of NGC 5728 and NGC 3393, there is energy radial dependence in the cross-cone \textit{and} cone directions, which could be explained if the ionization bicone holds some inclinations with respect to the host disk, such as the photons from the two regions interact with both the ISM at larger radii \textit{and} with the innermost dense clouds.\par 

\begin{figure}[htb!]
  \centering
\begin{minipage}[b]{\textwidth}
%\hspace*{-1.25cm} 
  \includegraphics[width=18.0cm]{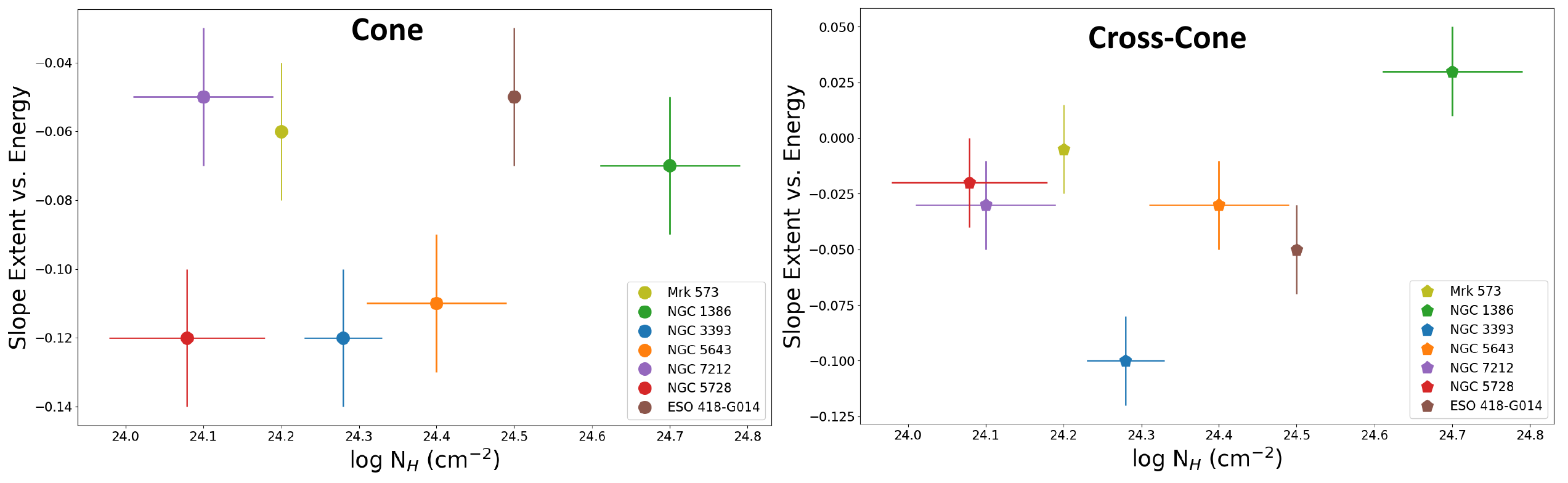}
 \end{minipage}\qquad\\
\begin{minipage}[b]{\textwidth}
\hspace*{-.14cm} 
  \includegraphics[width=18.0cm]{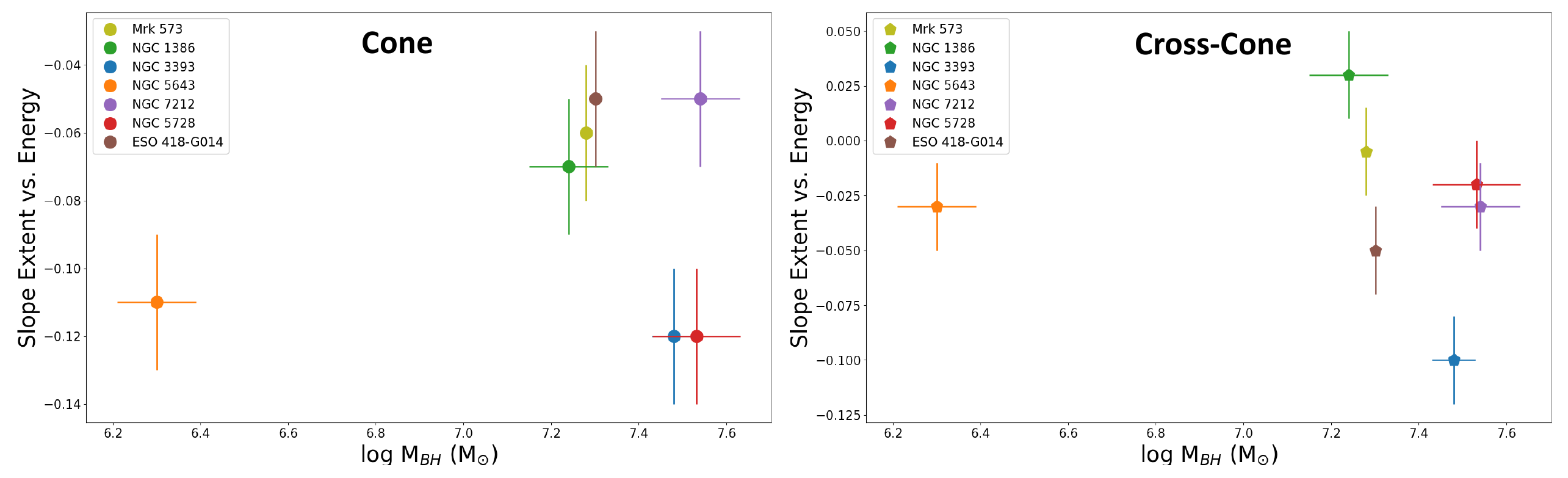}
 \end{minipage}\qquad\\ 
 \begin{minipage}[b]{\textwidth}
%\hspace*{-.05cm} 
  \includegraphics[width=18.0cm]{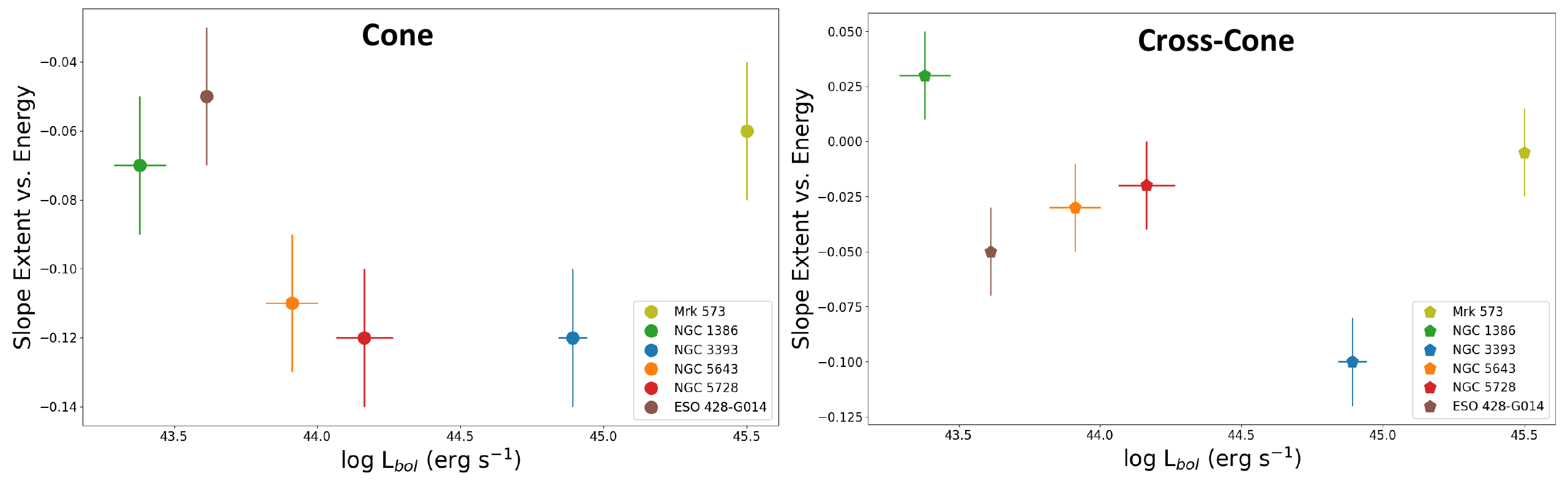}
 \end{minipage}\qquad\\
 \begin{minipage}[b]{\textwidth}
%\hspace*{-.06cm} 
  \includegraphics[width=18.0cm]{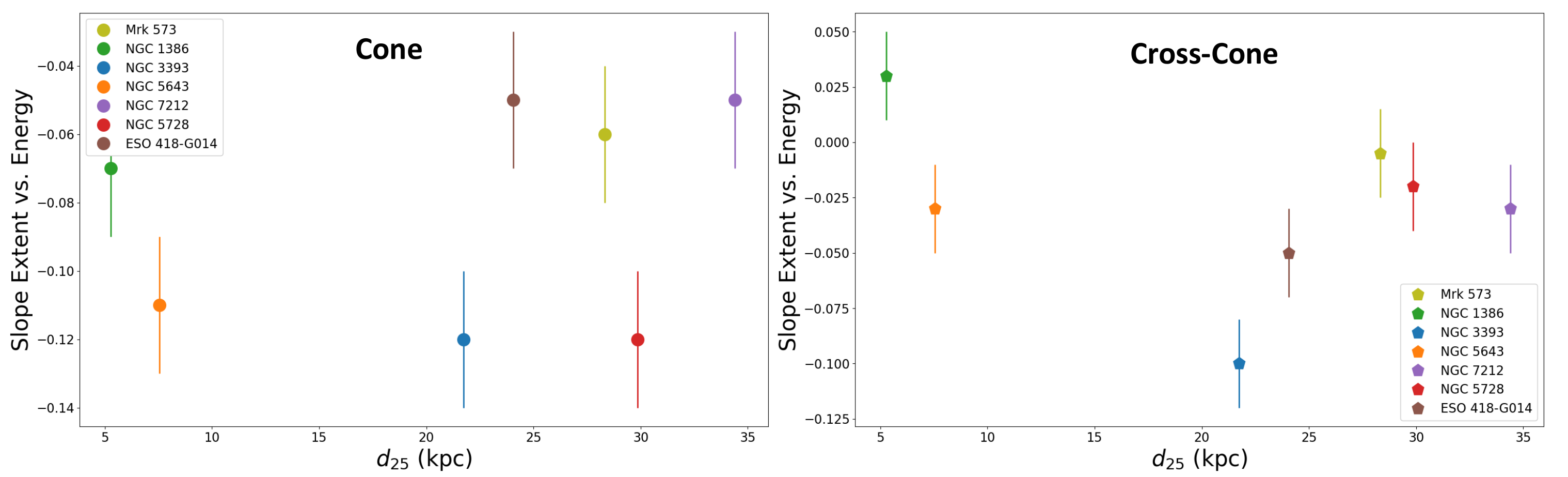}
 \end{minipage}\qquad\\
\caption{Best-fit slope of the normalized emission extent at 1\% of the surface brightness \textit{vs.} energy in the cone (left column) and cross-cone (right column) regions as a function of (top to bottom): column density, black hole mass, bolometric luminosity, and host galaxy diameter. Adapted from \citealt{fabbiano2018a} and \citealt{jones2021a}.}
 \label{fig:plot_prop} 
 \end{figure}

Following the approach adopted by \citet{jones2021a}, we plot the slope of the normalized extent-energy relation \textit{vs.} black hole and host galaxy properties, such as column density, black hole masses, bolometric luminosity, and host galaxy diameter (Fig. \ref{fig:plot_prop}). We find little to no correlation between the extent-energy slope and the obscuring column density, black hole mass, and galaxy diameter (Pearson coefficients show a "weak" correlation for cone and cross-cone directions). On the other hand, we find a tendency for the bolometric luminosity in the cross-cone direction, such that the less luminous sources tend to have the soft X-ray extent dominating the extent in the hard band (Pearson coefficient $\sim$ -0.65).

%We also measure the extent of the diffuse emission in the Fe K$\alpha$ line band (6.1-6.6 keV) (Figure \ref{fig:3-6kev}, bottom). In the cone direction, the emission is extended $\sim$ 5.0$''$ (1 kpc), but is consistent with the PSF in the cross-cone direction.

With our deep \textit{Chandra} observations of NGC 5728, we can also determine the relative amount of cross-cone to cone emission in different energy bands. In Table \ref{tab:ratio}, we list the ratio of the X-ray emission counts (cross-cone/cone), for different energy bands, from the 1.5$''$-8$''$ conical sectors. Regions containing off-nuclear point sources were excluded. Counts are derived by subtracting the normalized model PSF for each energy band. As in ESO 428-G014 \citep{fabbiano2018a}, the cross-cone emission is more abundant for energies $<$ 1.5 keV and fairly constant for energies $>$ 3 keV. \par

\begin{table}[htb!]
\footnotesize
\begin{center}
\caption{\textbf{Extended "Cross-Cone"/"Cone" Emission Ratios}}
\label{tab:ratio} 
\begin{tabular}{cccccccc}
\multicolumn{8}{c}{}\\

\hline
\multicolumn{1}{c}{}
&\multicolumn{1}{c}{0.3-3.0}
&\multicolumn{1}{c}{0.3-1.5}
& \multicolumn{1}{c}{1.5-3.0}
&\multicolumn{1}{c}{3.0-6.0}
&\multicolumn{1}{c}{3.0-4.0}
&\multicolumn{1}{c}{4.0-5.0}
&\multicolumn{1}{c}{5.0-6.0}

\\
Energy Range (keV) & --------------- & --------------- & --------------- & --------------- & --------------- & --------------- & --------------- \\
& Ratio (error) & Ratio (error) & Ratio (error) & Ratio (error)& Ratio (error) & Ratio (error) & Ratio (error) \\
\hline
& 0.16 (0.02)&0.20 (0.02)&0.10 (0.01)&0.05 (0.01)&0.04 (0.01)&0.01 (0.01)&0.03 (0.02)\\
\hline
\end{tabular}
\end{center}
\end{table}

%The presence of extended diffuse X-ray emission in the cross-cone direction does not agree with the standard AGN model proposed by \citet{antonucci1993a}, in which an optically thick, uniform torus, exists around the nucleus, allowing radiation to escape only along the bicone direction. The same phenomenon has been found in previous CT AGN studies \citep[e.g.,][]{wang2011c,fabbiano2018a, travascio2021a, fabbiano2022a}, and the authors suggested that the torus might not be perfectly obscuring, allowing some of the radiation to escape in the cross-cone direction. We will further discuss this possibility in Section \ref{sec:contribution}.

\subsection{Interpretation of the Physical Models}
\label{sec:interpretations}

The results of our spectral analysis in Section \ref{sec:results} show that in the central 1.5$''$ nuclear region, the best-fit models require at least 2 photoionization components to adequately reproduce emission features over the entire energy range. The addition of a third thermal component with k$T$=1.4 keV is also acceptable (Table \ref{tab:parameters_models} and Fig. \ref{fig:nuc_spec}), suggesting that shocks of $\sim$ 1000-1200 km/s could be occurring in the central 1.5$''$ (300 pc) nuclear region (assuming $v_{shock}^2$ = 16$k$ T$_{shock}$/3$\mu$; where $k$ is the Boltzmann constant, and $\mu$ is the mean molecular mass of a fully ionized gas; e.g., \citealt{fabbiano2018a}). The predominance of photoionization processes in the nucleus is supported by the fact that emission of highly ionized species, such as [Si~VI] (1.96$\mu$m), was detected in this object \citep{durre2018a} extending out to $\sim$ 300 pc in each direction from the AGN. These emission lines indicate photoionization either from the AGN accretion disk or by fast shocks. The presence of a one-sided radio jet, impacting the gas at about 200 pc from the nucleus \citep{durre2018a}, may be connected to these shocks, which would have originated from the interaction between the radio jet emission and the ISM. Since the fit statistics are very similar in the central region for different multi-component models (Table \ref{tab:parameters_models}), both photoionization-only and mixed models are statistically acceptable and probable.  \par 

%In the extended region (1.5$''$-8$''$ annulus, 300-1600 pc), photoionization only models are still preferred, but as for the nuclear fit, the addition of a third thermal component produces a fit that is also acceptable, with kT$\sim$1.2 keV. As discussed by \citet{durre2018a}, the X-rays observed in this region can be generated either by direct photoionization from the central source, inverse Compton scattering associated with the radio jet, or by fast shocks. The idea of a thermal component in the extended emission as an indicator of fast shocks is supported by the fact that the kinematic maps from \citealt{durre2019a} show that the X-ray gas in this area possesses velocities with the required magnitude, and are co-located with the detected [Si~VI] emission generated by shocks.\par

For the emission detected in the bicone, our model results show that 2- thermal components are preferred to fit the spectra from the two cone regions. The \textsc{apec} output temperatures for the best-fit models (k$T1$=0.6$\pm$0.15 and k$T2$=1.4$\pm$0.27, for the NW cone; k$T1$=0.8$\pm$0.11 and k$T2$=1.3$\pm$0.11, for the SE cone) suggest shocks of $\sim$ 700 km/s, 1100 km/s, for the NW cone, and $\sim$ 850 km/s and 1100 km/s, for the SE cone. The thermal components in the cone regions may come from the presence of a radio jet extending in the SE-NW direction or from the interaction between X-ray winds and the ISM \citep{trindadefalcao2021b}. In the latter case, X-ray outflows that originated close to the AGN entrain small clouds of gas in the ISM, accelerating them to high velocities. In \citealt{trindadefalcao2021b}, the authors detect filaments of [O~III] gas traveling at $\sim$ 600 km/s and show that shocks with an X-ray wind with velocities $\sim$ 1500 km/s could have caused this acceleration. These velocities are in the same order of magnitude as the shock velocities found in the bicone regions in NGC 5728. \par

As shown in Table \ref{tab:parameters_models}, the SE conical region has a higher model flux ($\sim$1.5x) compared to the NW region, consistent with the fact that the SE cone presents more extended emission than the NW cone in all energy bands. This can be related to the obscuration of the NW X-ray emission by being behind the disk of the galaxy and the star-forming ring \citep{durre2018a}. Overall, we find that the ionization mechanisms for NGC 5728 are primarily due to AGN activity, with some thermal contribution in the form of shocks in the cones, consistent with the line-ratio diagnostics from \citealt{durre2018a}.\par 

The photoionization parameters from our best-fit models with \textsc{cloudy}, i.e., U and NH, can be used to constrain and identify the presence of highly ionized outflows. Conditions in X-ray bicones are similar to those of warm absorbers (WAs) \citep[e.g.,][]{guainazzi2005a}. These typically have line-of-sight velocities that range in the 1000-2000 km/s, with NH $\sim$ 10$^{20}$ - 10$^{21}$ cm$^{-2}$ \citep[e.g.,][]{fischer2013a}, and have, in most cases, originated in the inner 100 pc \citep[e.g.,][]{arav2015a, krongold2007a}.\par

\begin{figure}[htb!]
  \centering

 \includegraphics[width=10cm]{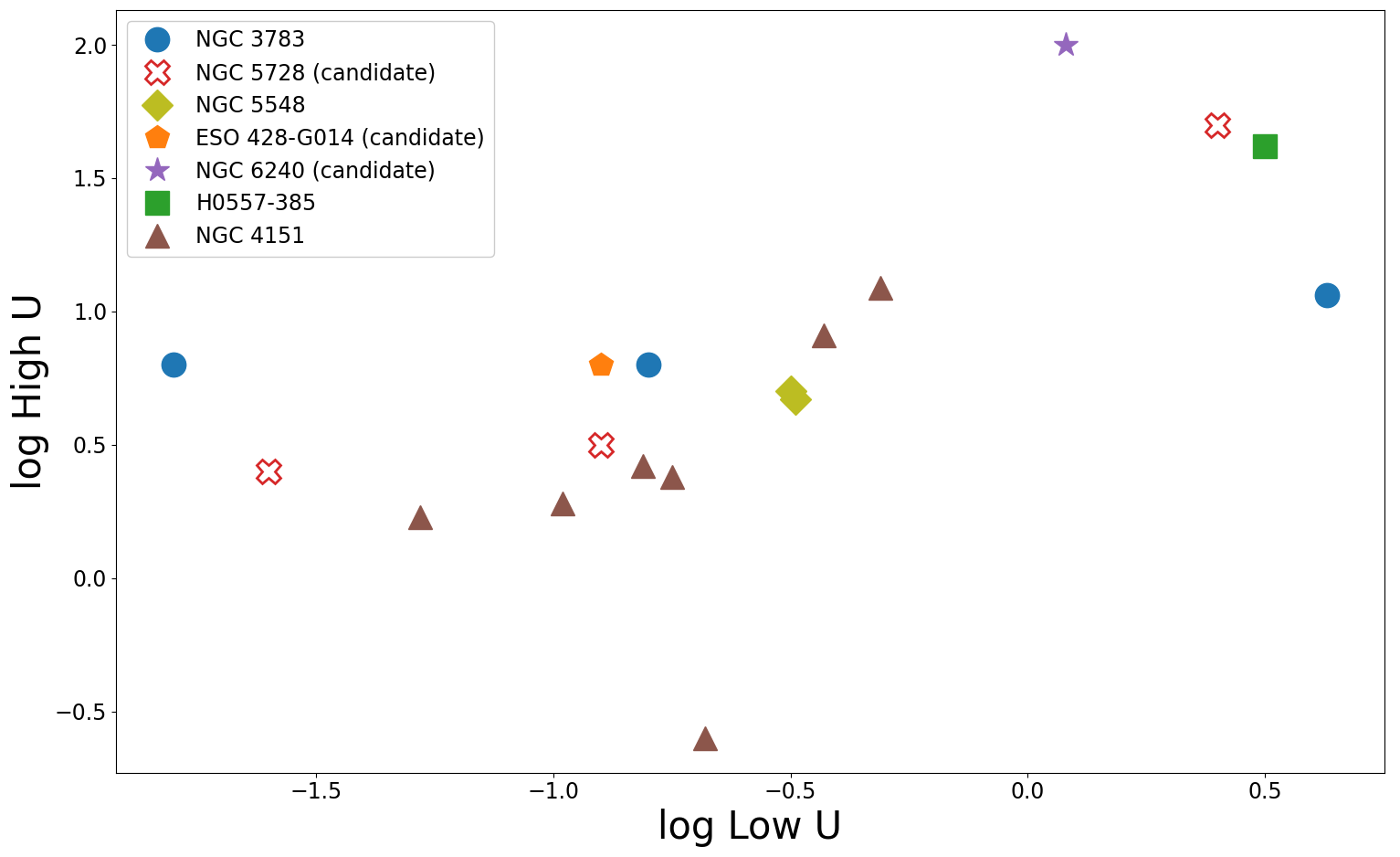}
\caption{Comparison between the low (x-axis) and high (y-axis) ionization parameters for WAs (and candidates) found in the literature, and our values for the WA candidate found in NGC 5728.
 The numbers for the best-fit parameters for the Seyfert galaxies in the plot were obtained from \citealt{andrade-velazquez2010a} (NGC 5548), \citealt{fabbiano2018a} (ESO 428-G014), \citealt{kaspi2002a} (NGC 3783), \citealt{krongold2003a} (NGC 3783), \citealt{paggi2022a} (NGC 6240), \citealt{ashton2006a} (H0557-385), \citealt{mizumoto2016a} (NGC 4051).}
 \label{fig:WA_param} 
 \end{figure}

From Table \ref{tab:parameters_models}, the photoionization parameters from the nuclear emission are log(U1)=0.4 and log(U2)=-1.6, consistent with what has been found in the literature for 2 phase (high and low ionization components) WA models. For instance, in \citealt{kaspi2002a}, the authors find that the modeled WAs for NGC 3783 have log(U1)=0.8 (high-velocity component) and log(U2)=-1.8 (low-velocity component). Similarly, in NGC 5548 \citet{andrade-velazquez2010a} find log(U1)=0.7 (high-velocity component) and log(U2)=-0.5 (low-velocity component). However, the column density of the nuclear low-velocity component is too high ($\sim$ 100x higher) compared to what is observed in these WAs. The NW bicone region, on the other hand, contains a high-velocity ($\sim$ 1200 km/s) thermal component and U and NH that are more consistent with WAs. In Fig. \ref{fig:WA_param}, we plot a comparison between the ionization parameters of 2 phase WAs (or candidates) found in the literature and the values found for NGC 5728 in the NW cone. \par

The identification and characterization of WA spectral components can be used to construct physical models of these outflowing, highly ionized gas. While we cannot affirm that WAs are present in NGC 5728, the NW cone region is a good candidate to host these high-velocity winds.

\subsection{Cone and Cross-Cone Contributions}
\label{sec:contribution}

As discussed in Sections \ref{sec:images} and \ref{sec:results}, we detect extended emission in the soft band, both in the cone and cross-cone directions. The NW and SE cones contribute $\sim$34\% of the total observed counts (in the broad-band, in a 8$''$ radius circle), with the NW cone contributing $\sim$15\% and the SE cone $\sim$19\%. For a 1.5$''$-8$''$ annulus region, the NW cone contributes $\sim$44\% and the SE cone $\sim$45\%, with the cross-cone contributing the remaining $\sim$11\% of the total observed counts in the 0.3-7 keV energy band in this region. Regarding the emission below 3 keV, the total contribution of the NW cone is $\sim$22\% of the total observed counts in a 8$''$ radius, while the SE cone contributes $\sim$33\%. For a 1.5$''$-8$''$ annulus region, the NW cone contribution is $\sim$46\%, and the SE cone contribution is $\sim$50\%. The lower contribution percentage of the NW cone compared to the SE cone might be related to the obscuration caused by the star-forming ring or the host galaxy disk to the NW region, as discussed. \par 

By comparing the extent of the X-ray emission at 1\% surface brightness with energy (Fig. \ref{fig:FWHM}), we find that both the cone and cross-cone emissions are more extended than the model PSF for energies below 3 keV. For the cross-cone direction, the soft X-rays extend on kpc scales. This emission is detected in a direction where the radiation from the AGN should be obscured by the torus, based on the AGN unification model \citep{antonucci1993a}. The origin of this emission is still unknown, but if it originated in the nuclear region, this radiation is likely produced by the corona and has interacted with the surrounding ISM, propagating to larger radii. \par 
%\citet{semena2019a} fit the \textit{NuSTAR} and \textit{Chandra} spectra of NGC 5728 with a torus model (MYTorus\footnote{http://mytorus.com/}) and found that the spectra best-fit parameters are NH=1.2$\times$10$^{24}$~cm$^{-2}$, with an energy cutoff fixed at 200 keV. \par 

The idea of a "porous", clumpy torus has been suggested by other authors before to explain this excess emission \citep[e.g.,][]{wang2011c, fabbiano2017a}, which would occur along the plane of the obscuring component. We can estimate the transmission in the cross-cone direction by considering the cone and cross-cone's solid angles (See Section \ref{sec:images}). Since the total bicone region volume is $\sim$2.5x larger than that of the cross-cone, the transmission in the cross-cone direction needs to be $\sim$ 2\% of that of the bicone direction. This is lower than the cross-cone transmission estimated for other CT AGN in the literature (e.g., ESO 428-G014 (\citealt{fabbiano2018a}), $\sim10\%$; NGC 7212 (\citealt{jones2020a}), $\sim16\%$).\par

Another possibility is that this emission in the cross-cone direction is related to the radio jet seen at 6cm, aligned with the NW-SE bicone emission \citep[e.g.,][]{durre2018a}. The presence of warm and hot emissions due to jet/cold-disk interactions have been modeled using relativistic hydrodynamical simulations \citep[e.g.,][]{mukherjee2018a}, and models have shown that gas with temperatures $\sim$10$^{7}$K may be ejected in the form of winds from the central regions on large scales, surrounding the cooler gas present in these regions. This is consistent with the temperatures found in our best-fit models for the NW and SE cone regions (kT$\sim$10$^{7}$~K$\sim$0.9 keV), and best alternative model for the central 1.5$''$ region.\par

Finally, a third possibility is that the origin of this excess extended emission is not related to nuclear activity but emission produced by hot ISM connected with the star-forming ring. \citet{durre2018a} presented the VLA 20cm radio large-scale map of NGC 5728 and showed that the radio emission exists in the nuclear region and also in two regions coinciding with two regions of star formation at the end of the galaxy bar. \par

%In the case of the cross-cone emission, recent simulations show that a hot cocoon of gas with T$\sim$10$^{8-9}$~K may surround the nuclear region due to jet-ISM interactions. As shown in Table \ref{tab:parameters_models}, we find a thermal component in the SE bicone region with higher temperatures (kT$\sim$1.3 keV) than in the central region (kT$\sim$0.96 keV), however, these temperatures are still not high enough to indicate the presence of these hot gas cocoons (kT$\sim$10$^{8}$K$\sim$ 7.0 keV).\par

%Finally, another possibility is that the gas in the nuclear and extended regions is being heated up by supernova activity in the star-forming ring. We will explore this possibility in a subsequent paper.

\section{Conclusions}
\label{sec:conclusions}

We analyze the deep \textit{Chandra} observation of the CT AGN NGC 5728 to study the kpc-scale diffuse X-ray emission, both spectrally and spatially, as a function of energy. In summary, our conclusions are as follows:\par

\textbf{1.} We detect a lower energy total extent of $\sim$ 4 kpc in the direction of the major axis of NGC 5728, which is also the direction of the ionization bicone \citep{durre2018a}. We find an energy-related trend in the size of the extended emission, with the softer emission ($<$3 keV) being more extended than the harder emission. The smaller extent of the hard continuum and Fe K$\alpha$ profiles imply that the optically thicker clouds responsible for this scattering may be relatively more prevalent closer to the nucleus. These clouds must not prevent soft ionizing X-rays from the AGN from escaping to larger radii to have photoionized ISM at larger distances. Therefore, these scattering clouds must be clumped and have higher density. This suggests that at smaller radii, there may be a larger population of molecular clouds to scatter the hard X-rays, as in the Milky Way \citep[e.g.,][]{nakanishi2006a};\par 

\textbf{2.} The diffuse emission is also extended in the cross-cone direction, where the AGN emission would be mostly obscured by the torus in the standard AGN model. The presence of hard X-rays detected on kpc scales suggests a non-uniform, clumpy torus structure \citep[e.g.,][]{nenkova2008a}, which allows the radiation to escape from the circumnuclear region to larger radii. If this is the case, we estimate that the transmission in the cross-cone direction is $\sim$2\% of the bicone direction. This is lower than the transmission found by \citet{fabbiano2018a} for ESO 428-G014 ($\sim$10\%), and by \citet{jones2020a} for NGC 7212 ($\sim$16\%); \par 

\textbf{3.} We show that the extent as a function of energy at 1\% of the surface brightness in the cone region of NGC 5728 exhibits a steeper relationship than the cross-cone region. This may be explained by an inclination effect, in which the orientation of the ionization bicone with respect to the host disk molecular clouds impacts the extent to which the soft X-rays propagate through the ISM. We also compare the energy-extent slopes to galaxy and black hole properties and find for the cross-cone that the lower luminosity targets tend to have the soft X-rays extending farther than the hard X-rays. \par

\textbf{4.} We extract the spectrum from four different regions: a 1.5$''$ (300 pc) nuclear region, two 1.5$''$-8$''$ (300 - 1600 pc) conical sectors, namely SE and NW bicone, and one 2$''$-8$''$ (400 - 1600 pc) conical region, namely cross-cone. Using a nuclear \textsc{pexrav} + \textsc{xszcutoffpl} model to represent the reprocessed emission (for the nuclear fit only) + additional power-law to fit the soft emission and a set of Gaussian lines \cite[e.g.][]{levenson2006a}, we separately analyzed the emission in these regions. The model fits suggest the presence of blended O~VII, O~VIII, Ne~IX, Ne~X, and several Fe lines at energies $<$1.2 keV in these regions. We also detect more isolated emission from lines that can be identified as Mg (XI, XII), and Si, which is consistent with the findings in other AGN \citep[e.g.,][]{koss2015a, fabbiano2018a,travascio2021a};\par 

\textbf{5.} As reported by \citet{fabbiano2017a} for the soft component, all these spectral components are associated with large-scale spatial components, although point-like nuclear emission can also be seen at the higher energies and in the Fe K$\alpha$ line. The observed continuum emission in the 1.5$''$-8$''$ annulus accounts for $\sim$39\% of the total observed continuum in the 0.3-7 keV band. In the hard band (3.0-7.0 keV), 19\% of the continuum is observed in the 1.5$''$-8$''$ annulus; \par 

\textbf{6.} We detect 4 emission features that are not typical of CT AGN. The emission lines at $\sim$2.8 keV, and $\sim$3.5 keV were tentatively identified as Ar~K$\alpha$ (2.9 keV), Ar~XVII (3.7 keV), and Ca~K$\alpha$ (3.7 keV). The emission features to the red ($\sim$6.1 keV) and to the blue ($\sim$6.8 keV) of the 6.4 Fe K$\alpha$ line, are observed in the spectra extracted from all three regions, but could not be entirely identified;\par

\textbf{7.} We fit the spectra with photoionization and/or thermal models. In the inner 1.5$''$ circumnuclear region, the spectrum is best fit with a minimum of 2-photoionization components, although adding a third thermal component is also acceptable. For the conical regions, the NW and SE cones are best fit with at least 1- phototionization + 1- thermal components, although 2- photoionization + 2- thermal components are preferred for the NW conical region;\par

\textbf{8.} Physical interpretation of our fitting results suggests that the best-fit thermal components for NGC 5728 can be explained by ISM interactions, shocks, or supernova remnants/star formation from the nearby star-forming ring. Specifically:\par 
- In the inner 1.5$''$ central region, the alternative model fit suggests temperatures of $\sim$1.4 keV, resulting in $\sim$1000-1100 km/s shocks. For the extended emission (1.5$''$-8$''$ conical regions), the thermal components have similar temperatures compared to those of the nuclear region, $\sim$ 1.4 keV and 1.3 keV, corresponding to shocks $\sim$1100-1200 km/s. \par
-The lower temperature components in the cone regions (kT$\sim$ 0.6-0.8) can also be explained by jet-ISM interactions. In this case, cool gas in the nucleus is surrounded by warm gas in the central region, with the latter being ejected from the nucleus on large scales in the form of winds.\par
-The possibility of supernova heating and/or heating from star-forming regions is not ruled out. We will further explore this possibility in a subsequent paper;\par

\textbf{9.} We find that our models' ionization parameters for the inner 1.5$''$ emission are consistent with those found in warm absorbers (WAs). However, the typical column densities values for WAs are more consistent with the parameters obtained for the preferred model in the NW cone region rather than in the very central source;\par

\medskip
\medskip

This paper demonstrates how deep \textit{Chandra} observations can produce surprising results on "standard" sources, recovering crucial information about the AGN and surrounding ISM. With statistically significant spectral and spatial information, these observations have uncovered large-scale diffuse X-ray emission and have allowed us an opportunity to study and test the AGN standard model, providing new insights into the physics of AGN and AGN feedback.

\begin{acknowledgments}
The authors thank the anonymous referee for helpful comments that
improved the clarity of this paper. We thank Margarita Karovska and Steve Kraemer for their valuable comments on the manuscript. This work was partially supported by NASA contract NAS8-03060 (CXC) and the \textit{Chandra} Guest Observer program grant GO0-21094X (PI: Fabbiano). The NASA ADS bibliography service was used in this work. 
This research has made use of the NASA/IPAC Extragalactic Database (NED), which is operated by the Jet Propulsion Laboratory, California Institute of Technology, under contract with the National Aeronautics and Space Administration. This work used the photoionization code \textsc{cloudy}, and we thank Gary Ferland and associates for the maintenance and development of the software. 
\end{acknowledgments}

\bibliographystyle{aasjournal}
\bibliography{anna_bibliography}

\appendix

\section{Alternative Fitting Models}

\begin{figure*} [h!]
    \centering
    \includegraphics[width=18cm]{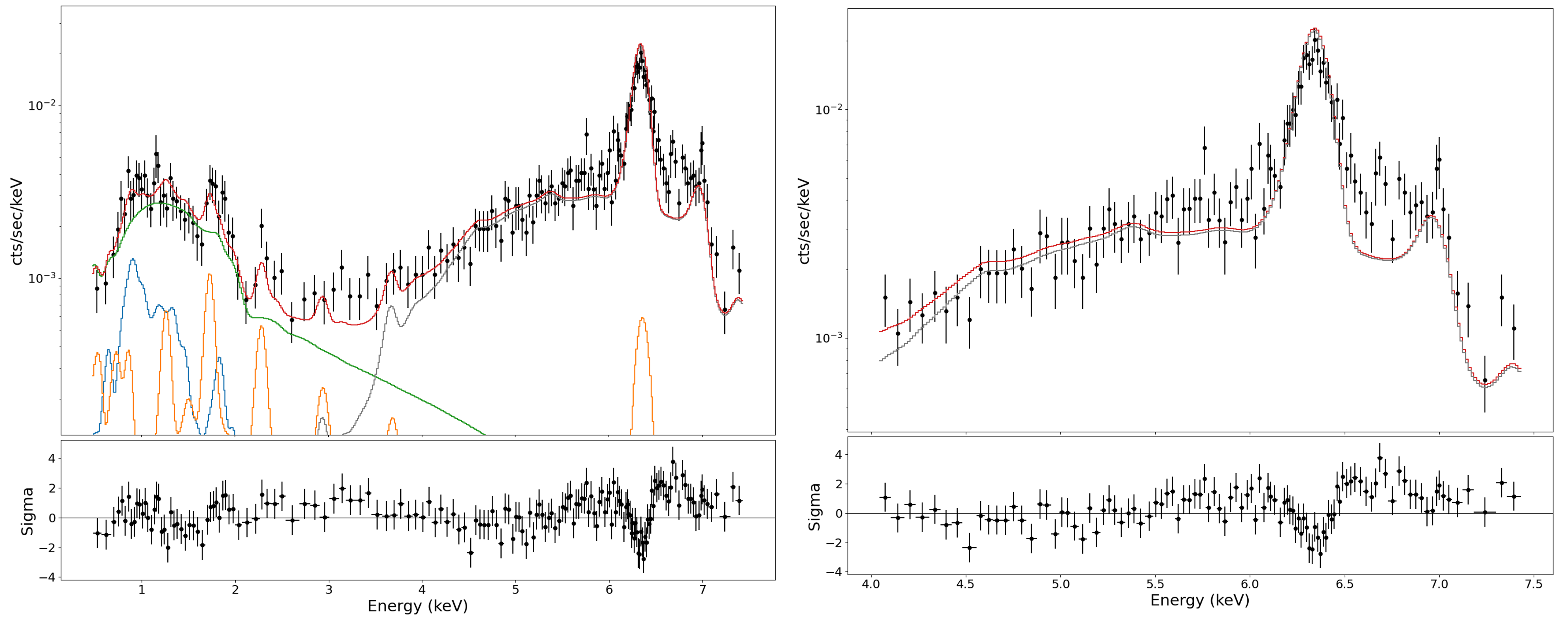}
    \caption{\textbf{Left:} Full-band (0.3-8 keV) spectrum from the inner 1.5$''$-radius circle with the best-fit (top) and best-fit residuals (bottom) for a 2-photoionization component model. The hard X-ray spectrum was fit with \textsc{borus02} with $\Gamma$=2.1 (in gray). As shown, the excess emission seen to the left and right of the neutral Fe K$\alpha$ still appears as residuals at $\sim$6 and 7 keV. \textbf{Right:} Hard (4-8 keV) spectrum from the inner 1.5$''$-radius circle with the best-fit \textsc{borus02} model (top) and best-fit residuals (bottom), shown in more detail.}
    \label{fig:borus}
\end{figure*}

We fit the hard nuclear spectrum with \textsc{borus02} \citep{balokovic2018a}, as an alternative model for \textsc{pexrav+xszcutoffpl}. The reprocessed component represented by this model is produced by the obscuring material near the SMBH, and it includes the scattered component and fluorescent lines (see the \textsc{borus02} website\footnote{http://www.astro.caltech.edu/~mislavb/download/} for details). We fixed the energy cutoff at 200 keV, and assumed the cosine of the angle at which the observer looks at the plane layer to be cos($\theta$)=0.5 \citep[e.g.,][]{semena2019a}. The results of this analysis are shown in Fig. \ref{fig:borus}. The best-fit model for this torus component returns a photon index $\Gamma$=2.1$\pm$0.3, and hydrogen column density component logNH=23.8$\pm$0.1. As shown, excess emission redward and blueward of the neutral Fe K$\alpha$ line is still observed in the bottom panels, near 6 and 7 keV.

\section{Emission Lines Identification}
\begin{longtable}{|p{3.5cm}p{4cm}p{3.5cm}p{3.5cm}p{1cm}|}
\caption{\textbf{Spectral Fitting Results for the Extracted Spectra}}
\label{tab:parameters_pheno} \\
\hline
\textbf{Region} & \textbf{Counts (error)} & \textbf{Norm. Po. Law} & \textbf{Model Energy Flux} & \textbf{$\chi ^{2}$}\\
& (0.3-7.0 keV) & (ph~cm$^{-2}$~s$^{-1}$)&  (0.3-7.0~keV; in erg/cm$^{2}$/s) & \\

\hline
 \textbf{(1.5$''$ circle)} & 4577 (67) & 4.0$\times$ 10$^{-6}$ & 2.5$\times$10$^{-12}$ & 0.69\\
 \textbf{(1.5$''$ - 8 $''$ NW Bicone)} & 1091 (28)& 6.0$\times$ 10$^{-6}$& 7.2$\times$10$^{-14}$ & 0.27\\
 \textbf{(1.5$''$ - 8 $''$ SE Bicone)} &1383 (32) &6.0$\times$ 10$^{-6}$ & 1.1$\times$10$^{-13}$ & 0.85 \\
 \hline
 
 \underline{\textbf{Emission Lines:}} & & & &  \\
 Observed Energy (keV) & Flux (10$^{-6}$ ph~cm$^{-2}$~s $^{-}$) & Identification (E$_{\rm lab}$ keV)$^{a}$ & & \\
 (1.5$''$ circle)& (1.5$''$ circle)& & & \\
 (1.5$''$ - 8 $''$ NW Bicone)& (1.5$''$ - 8 $''$ NW Bicone)& & & \\
 (1.5$''$ - 8 $''$ SE Bicone)&(1.5$''$ - 8 $''$ SE Bicone)& & & \\
\hline

- &- & *O~VII (0.57)  & &\\ 
-& -&   & &\\
0.46$\pm$0.08 &49.1$\pm$18.5  & & &\\
\hline 

-& -  & & &\\
0.73$\pm$0.05& 3.4$\pm$1.2 &*Fe~XVII (0.83) & &\\
0.76$\pm$0.01&5.0$\pm$0.01   & & &\\
\hline

0.82$\pm$0.01 & 4.83$\pm$0.72  & *O~VIII (0.65),& &\\
 -   & - & *Fe~XVII (0.83)& &\\
 -  & - & & & \\
\hline

-  & -  &*Ne~IX (0.90), & &\\
 - &- &  *Fe~XIX (0.92) & &\\
  0.90$\pm$0.02& 7.40$\pm$1.16&  & & \\
\hline

0.98$\pm$ 0.01 & 2.41$\pm$0.3  &*Fe~XXI (1.01), & &\\
0.99$\pm$0.07& 2.13$\pm$0.73  &*Ne~X (1.02)  & &\\
-& -  & & &\\
\hline

1.17$\pm$ 0.01 & 1.35$\pm$0.21  & *Fe~XXIV (1.13, 1.17) & &\\
-&- &   & &\\
1.18$\pm$0.03& 1.29$\pm$0.36&   & &\\
\hline

1.33$\pm$0.01 & 1.05$\pm$0.28  & Mg~XI (1.33)& &\\
-& -  & & &\\
-& -  & & &\\
\hline

1.75$\pm$0.01 & 0.58$\pm$0.12 & Si~XIII (1.84)  & &\\
1.82$\pm$0.05 & 0.43$\pm$0.15 &   & &\\
1.79$\pm$0.02 &0.38$\pm$0.14 &   & &\\
\hline

1.92$\pm$0.02 & 0.43$\pm$0.11   & Si~XIII (1.86)& &\\
 - & -&   & &\\
 - &- &   & &\\
\hline

2.32$\pm$0.02 & 0.49$\pm$0.14 & S~K$\alpha$ (2.31), & &\\ 
2.32$\pm$0.06  &0.42$\pm$0.18 & S~XV (2.43) & &\\
2.35$\pm$0.03  & 0.37$\pm$0.15& & &\\
\hline

 - &- & Ar K$\alpha$ (2.96)$^{\bigstar}$ & &\\
2.76$\pm$0.08 &0.11$\pm$0.09   & & &\\
2.95$\pm$0.06  & 0.15$\pm$0.08  & & &\\
\hline

3.14$\pm$0.05 & 0.17$\pm$0.09  & Ar~XVII (3.69)$^{\bigstar}$,  & &\\
3.54$\pm$0.11  & 0.39$\pm$0.16   &Ca~K$\alpha$ (3.69)$^{\bigstar}$, & &\\
3.61$\pm$0.06  &0.18$\pm$0.11   &blend of line & &\\
\hline

 - &- &Ca~K$\alpha$ (3.69)$^{\bigstar}$,  & &\\
4.65$\pm$0.08 & 0.26$\pm$0.12 & blend of lines   & &\\
4.49$\pm$0.17  &0.49$\pm$0.18   & & &\\
\hline

-& -  &Fe~K$\alpha$ "Red Wing"$^{\bigstar}$ & &\\
6.30$\pm$0.12 & 1.73$\pm$0.39  & & &\\
-  &-   & & &\\
\hline

6.40$\pm$0.01 & 18.02$\pm$0.91  & Fe~K$\alpha$ (6.44 keV) & &\\
6.45$\pm$0.04  &0.69$\pm$0.26   & & &\\
6.33$\pm$0.04  &0.83$\pm$0.23   & & &\\
\hline

\end{longtable}
\noindent$^{a}$ Energies from NIST; \citealt{koss2015a, maksym2019a, travascio2021a}.\\
* Lines blended in the ASCIS-S spectrum $<$1.3 keV. These are tentative identifications\\
$\bigstar$ These lines do not fit into our current understanding of AGN emission.

\end{document}